\documentclass[10pt,journal,compsoc]{IEEEtran}
\ifCLASSOPTIONcompsoc
  \usepackage[nocompress]{cite}
\else
  \usepackage{cite}
\fi

\ifCLASSINFOpdf
   \usepackage[pdftex]{graphicx}
\else
\fi

\ifCLASSOPTIONcompsoc
  \usepackage[caption=false,font=footnotesize,labelfont=sf,textfont=sf]{subfig}
\else
  \usepackage[caption=false,font=footnotesize]{subfig}
\fi
\usepackage[linesnumbered,ruled,vlined,noend]{algorithm2e}
\usepackage{caption}
\usepackage{tikz}
\usepackage{mathtools}
\DeclarePairedDelimiter\ceil{\lceil}{\rceil}
\DeclarePairedDelimiter\floor{\lfloor}{\rfloor}
\usepackage{enumitem,,lipsum}
\usepackage{amsmath}
\usepackage{amssymb}
\usepackage{enumitem}

\newtheorem{theorem}{Theorem}
\newtheorem{example}{Example}
\newtheorem{observation}{Observation}
\newtheorem{lemma}{Lemma}

\begin{document}
\title{Index-based Solutions for\\ Efficient Density Peak Clustering}

\author{Zafaryab~Rasool,
        Rui~Zhou,~\IEEEmembership{Member,~IEEE,}
        Lu~Chen,
        Chengfei Liu,~\IEEEmembership{Member,~IEEE,}
        and~Jiajie~Xu% <-this % stops a space
\IEEEcompsocitemizethanks{\IEEEcompsocthanksitem Z. Rasool, R. Zhou, L. Chen, C. Liu are with the Department
of Computer Science and
Software Engineering, Swinburne University of Technology, Melbourne, Australia. \protect\\
E-mail: \{zrasool, rzhou, luchen, cliu\}@swin.edu.au
\IEEEcompsocthanksitem J. Xu is with School of Computer
Science and Technology, Soochow University, Suzhou, China.
Email: {xujj}@suda.edu.cn.}% <-this % stops an unwanted space
}

\IEEEtitleabstractindextext{%
\begin{abstract}
Density Peak Clustering (DPC), a popular density-based clustering approach, has received considerable attention from the research community primarily due to its simplicity and fewer-parameter requirement. However, the resultant clusters obtained using DPC are influenced by the sensitive parameter $d_c$, which depends on data distribution and requirements of different users. Besides, the original DPC algorithm requires visiting a large number of objects, making it slow.
To this end, this paper investigates index-based solutions for DPC.
Specifically, we propose two list-based index methods viz. (i) a simple List Index, and (ii) an advanced Cumulative Histogram Index. Efficient query algorithms are proposed for these indices which significantly avoids irrelevant comparisons at the cost of space. For memory-constrained systems, we further introduce an approximate solution to the above indices which allows substantial reduction in the space cost, provided that slight inaccuracies are admissible. 
Furthermore, owing to considerably lower memory requirements of existing tree-based index structures, we also present effective pruning techniques and efficient query algorithms to support DPC using the popular Quadtree Index and R-tree Index. Finally, we practically evaluate all the above indices and present the findings and results, obtained from a set of extensive experiments on six synthetic and real datasets. The experimental insights obtained can help to guide in selecting a befitting index.
\end{abstract}

}

\maketitle

\IEEEdisplaynontitleabstractindextext

\IEEEpeerreviewmaketitle

\IEEEraisesectionheading{\section{Introduction}\label{sec:introduction}}

\IEEEPARstart{C}{lusters} reflect a potential relationship among different entities of data. 
This data can be sourced from a wide range of domains like market research, social network analysis, spatial data analysis, pattern recognition, etc. 
Many clustering algorithms have been developed in the last few decades in response to the proliferating demands across industries and organizations, which help them make operational and strategic decisions. 

Clustering inherently builds on the notion of similarity. However,
the notion of similarity is different in different clustering algorithms. Moreover, these clustering algorithms have different mechanisms to cluster a set of data objects. Among them, density-based clustering algorithms are popular, which find subsets of objects in ``dense regions" separated by not-so-dense regions, where each subset represents a cluster. 
In this paper, our focal point will be Density Peak Clustering (DPC), a popular approach towards obtaining density-based clusters, proposed by Rodrigues and Laio \cite{rodriguez2014clustering}. To highlight the significance of DPC, we first provide a comparison between \textit{density based} vs. \textit{non-density based} algorithms, and then the differences between DPC and DBSCAN \cite{ester1996density}, another popular density-based algorithm, are discussed.

(i) \textit{Density based vs. non-density based}: Density-based clustering algorithms have several advantages over non-density based algorithms, such as centroid-based and connectivity-based clustering algorithms. Density-based methods can find arbitrary-shaped clusters and do not need the number of clusters as input, unlike centroid-based clustering algorithms such as k-means. Moreover, they also identify outliers from the objects. The only drawback is that they require more execution time.
Connectivity-based methods (for e.g., \cite{king1967step}, \cite{sneath1962numerical}) organize objects in a hierarchy based on some similarity, but do not produce a unique partition.

(ii) \textit{DPC vs. DBSCAN}: The idea of DPC is similar to DBSCAN. DBSCAN identifies core objects first, which are then connected, if they are within distance $\epsilon$. The non-core objects are connected to the core objects nearby. If there is no nearby core object, a non-core object is treated as an outlier. Thus, a cluster in DBSCAN is recognized as a connected component of core and non-core objects. 
DPC directly links each object to the nearest object in a higher density region until the last such object (called a peak) is reached, forming a tree structure for each cluster. 
The outlier is distinguished as an object in a low density area and with a large distance to any other higher density object.
Thus, a cluster in DPC can be represented as a tree with the cluster center as the root. 
Both, DBSCAN and DPC, require a cut-off distance to determine the density of the area an object is located. DBSCAN also uses an additional local density parameter, i.e. minimum required points within the cut-off distance, to distinguish a core object and a non-core object.

\begin{figure*}[t]
\centering
\subfloat[$d_c=0.001$]{\includegraphics[width=1.7in, height=1.4in]{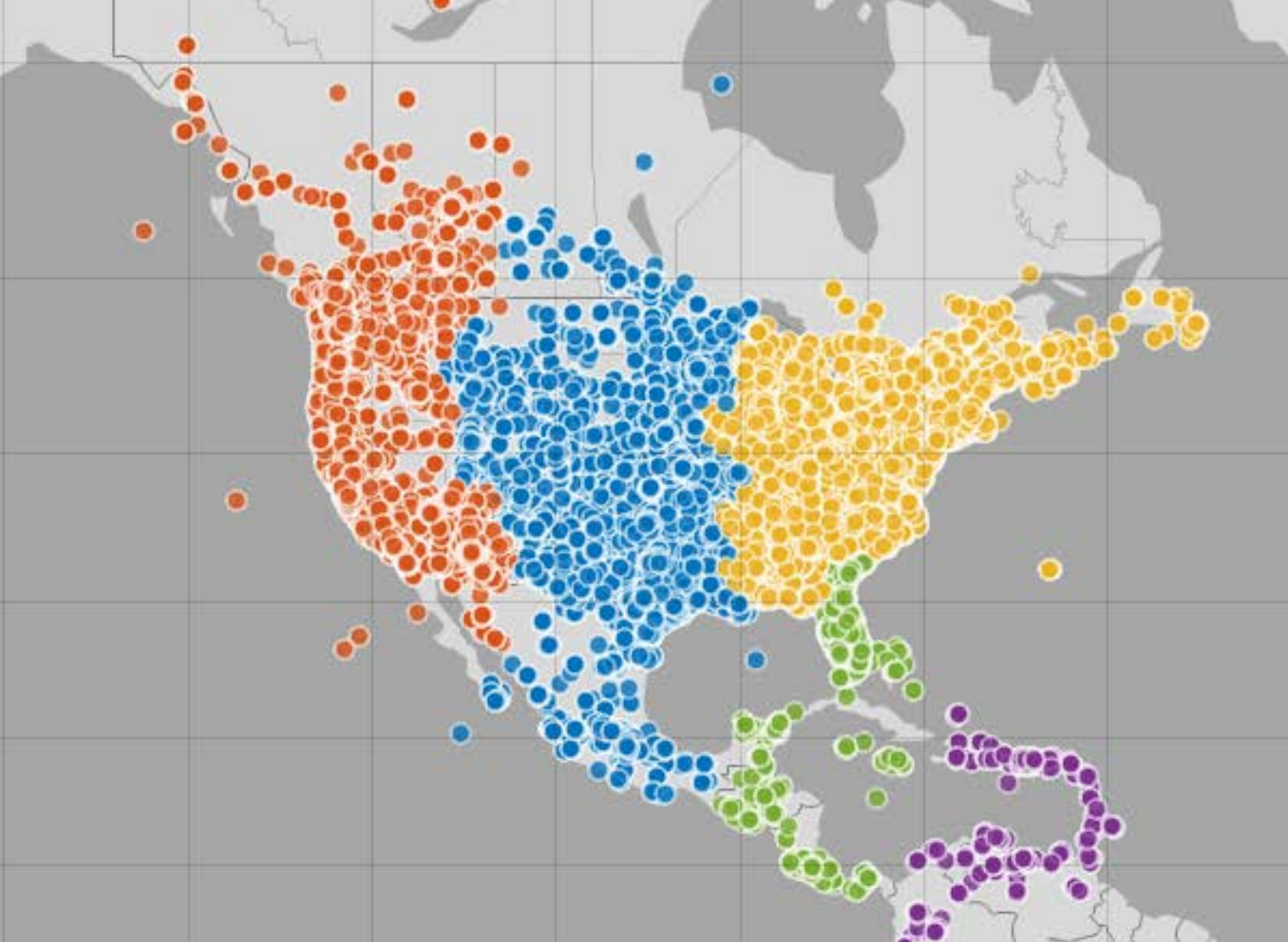}}
\label{fig:dpp1}
\hspace{0pt}
\centering
\subfloat[$d_c=0.01$]{\includegraphics[width=1.7in, height=1.4in]{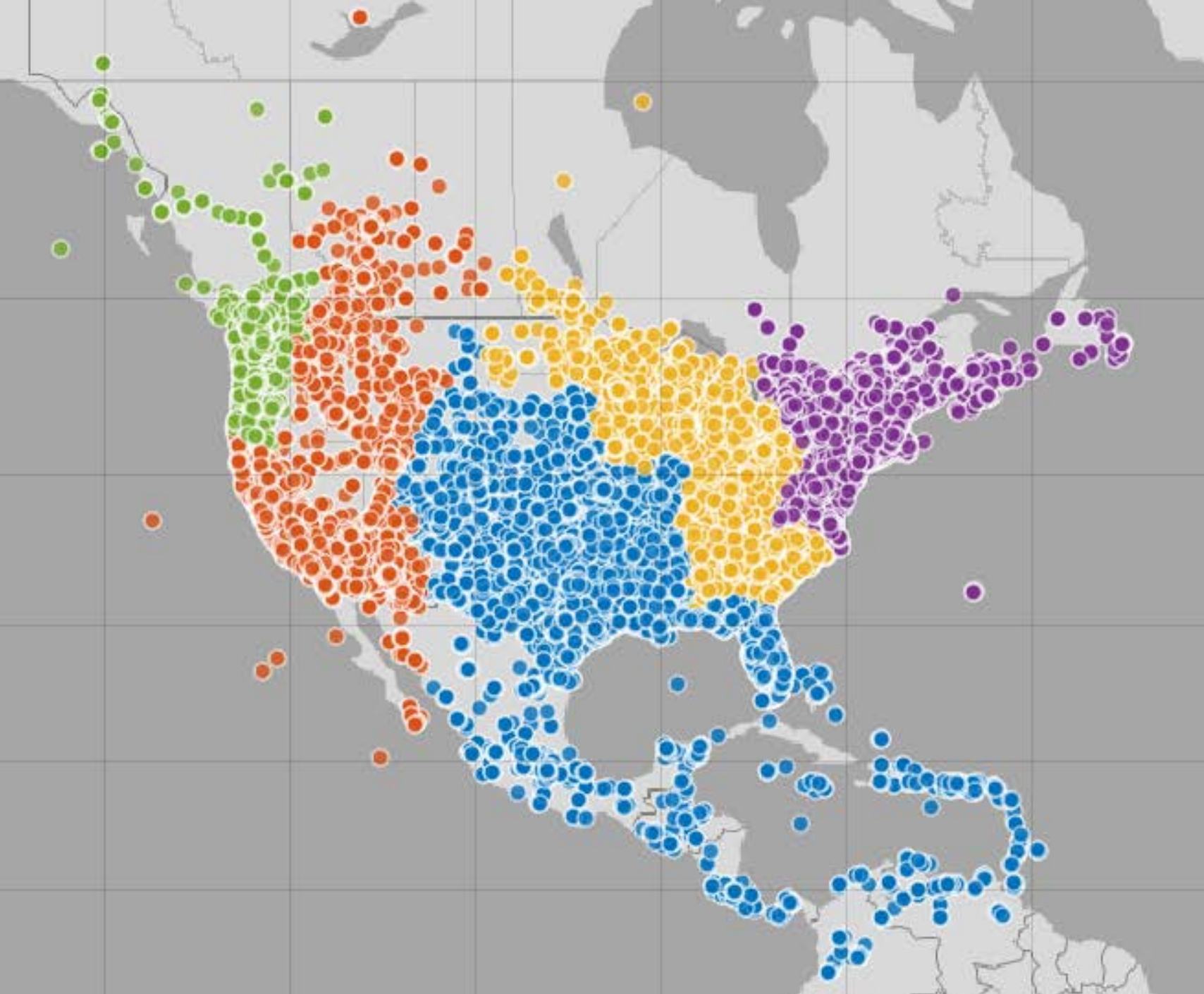}}
\label{fig:dpp2}
\hspace{0pt}
\centering
\subfloat[$d_c=1.0$]{\includegraphics[width=1.7in, height=1.4in]{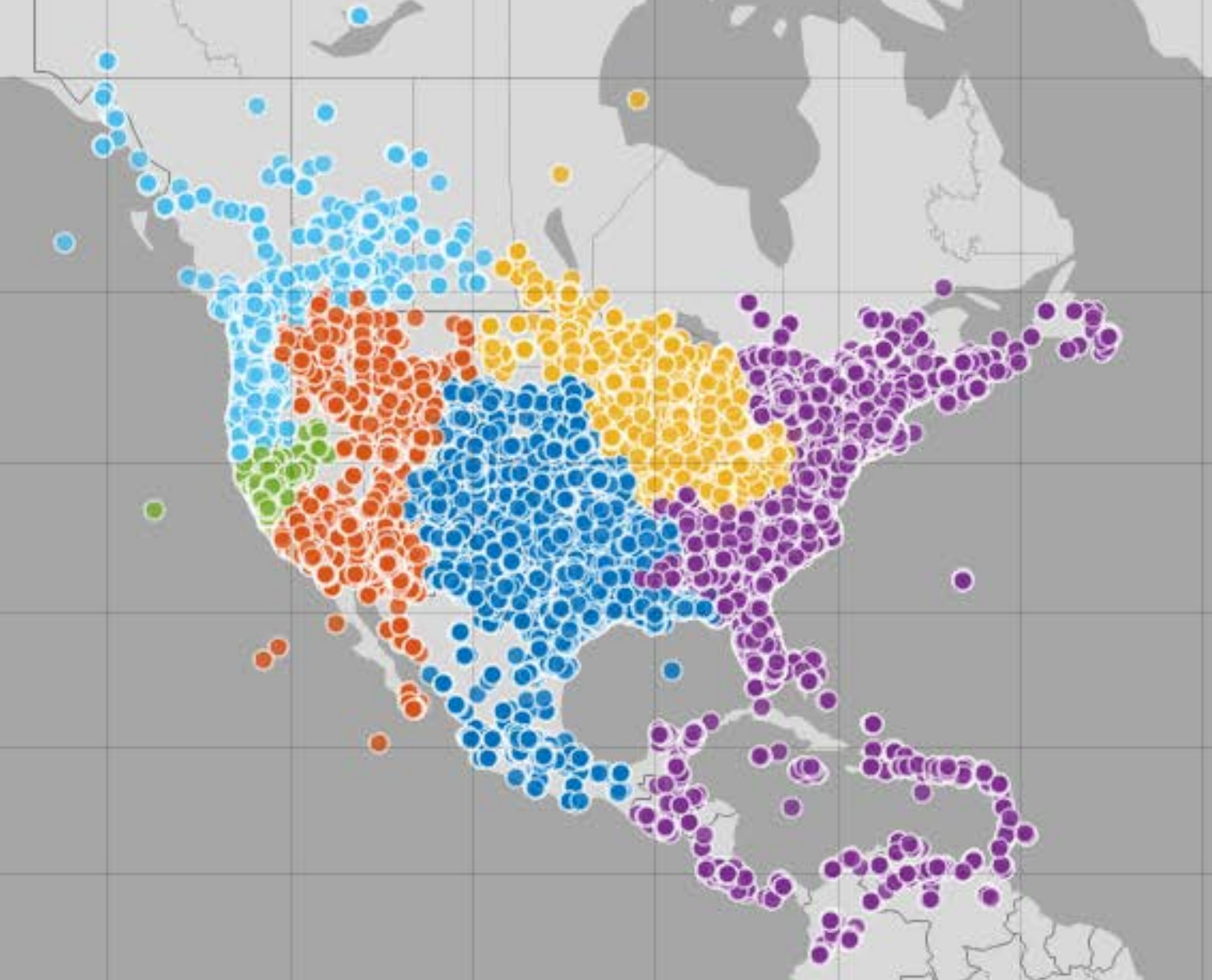}}
\label{fig:dpp3}
\hspace{0pt}
\centering
\subfloat[$d_c=10.0$]{\includegraphics[width=1.7in, height=1.4in]{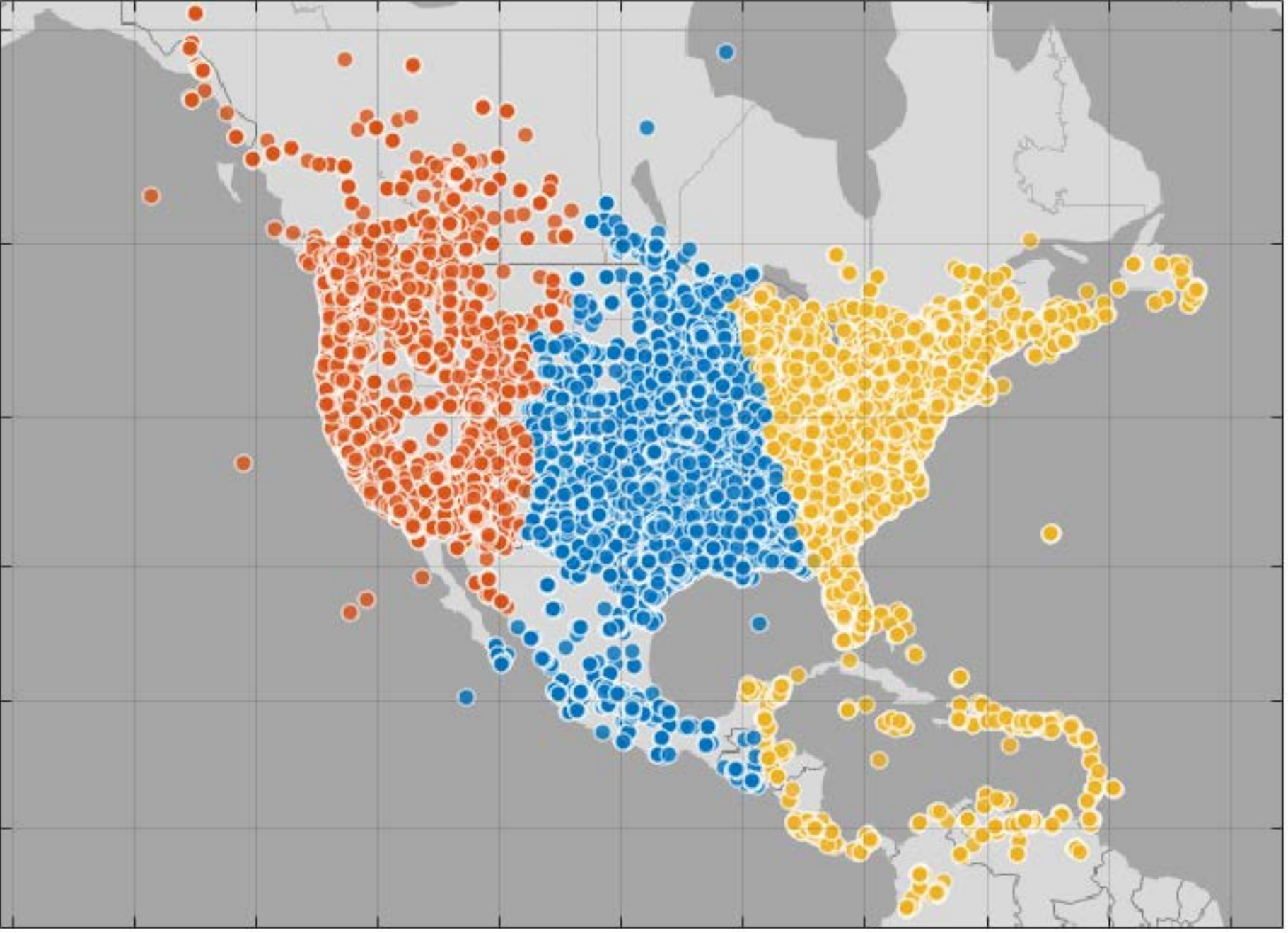}}
\label{fig:dpp3}
\caption{Different Clusters for different choices of $d_c$}
\label{fig:dptest1b}
\vspace{-8pt}
\end{figure*}

DPC aims to identify cluster centers among the set of objects, and for each object, it defines two quantities: (i) \textit{local density} $\rho$ and (ii) \textit{dependent distance} $\delta$. Given a set of objects P and parameter $d_c$, the \textit{local density} $\rho$ of an object $p \in P$ denotes the number of objects within distance $d_c$ from \textit{p}. The \textit{dependent distance} $\delta$ for an object $p \in P$ is the distance from $p$ to its nearest higher local density neighbor. Based on these, cluster centers are selected as objects having high $\rho$ and large $\delta$ values. Once the cluster centers are determined, the remaining objects are assigned to the clusters containing their nearest higher density neighbors. DPC has been employed by researchers to solve problems in domains, such as time-series \cite{begum2015accelerating}, neuroscience \cite{kobak2016demixed}, geoscience \cite{sun2015exemplar}, biology \cite{zamuner2015efficient}, computer vision \cite{dean2015high}, etc.

\medskip\noindent However, there are two major issues:
\begin{itemize}[leftmargin=*]
    \setlength\itemsep{0.0em}
    \item \textit{Parameter Selection.} DPC suffers from the parameter setting problem since its clustering results are heavily influenced by $d_c$. The selection of parameter $d_c$ is dependent on data distribution. Moreover, a user may have various requirements for clustering and may need to test several $d_c$ to obtain the desired clustering. This gets worse when different users may want to try different $d_c$.
    Figure~\ref{fig:dptest1b} illustrates that setting different $d_c$ will produce different clustering results based on the real-world dataset Gowalla containing user check-ins for the area around US and Caribbean. 
    
    \item \textit{Query cost.} Given a $d_c$, the original DPC algorithm needs to determine pair-wise distances and compute the two quantities $\rho$ and $\delta$ for each object. As this has high time cost for large datasets, running the DPC algorithm for different values of $d_c$ further exacerbates the problem.  
    
\end{itemize}

Motivated by the above issues, this paper aims to investigate employing index to speed up DPC for a given $d_c$. As such, the whole clustering process which probably involves trying many $d_c$ can be substantially shortened.
The importance of such a solution is impelled more by the fact that computing local density $\rho$ and dependent distance $\delta$ for each object are two expensive operations, and requires visiting a large number of objects.

%DPC involves expensive computation of two quantities for each object.
%The importance of such a solution is impelled more by the fact that there are two expensive queries for each object. 

To this end, we propose two list-based indices namely List Index and Cumulative Histogram (CH) Index. The List Index efficaciously captures the neighborhood of an object which facilitates fast access to close neighbors, bypassing the need to compare with most of the other objects. Based on List Index, efficient query algorithms are proposed to compute $\rho$ and $\delta$ for any $d_c$. However, for larger datasets, the computation of $\rho$ using List Index is still expensive. Therefore, the advanced CH Index is proposed which enables fast computation of $\rho$. Both of these indices have large memory requirements which may not be suitable for memory-constrained systems. Towards this, we present an approximate solution which can reduce the space cost with slight loss in accuracy.

Furthermore, we also study tree-based indices that are
capable of providing fast data access and have lower memory requirements. They can provide
efficient support to queries like range and nearest neighbor (\textit{NN}) search. 
Range search can be easily adapted to compute $\rho$, however,  a large number of search operations are still needed. Besides, the computation of $\delta$ is different from \textit{NN} search and requires a tailored algorithm. 
 In this context, we provide enhanced $\rho$ and $\delta$ queries based on effective pruning techniques using the popular Quadtree and R-tree indices for DPC.

\noindent
The key contributions of this paper are as follows:
\begin{itemize}[leftmargin=*]
\setlength\itemsep{0.0em}
\item We raise the importance of studying index for density peak clustering, supported by the fact that clustering results could be different using different distance settings. We aim to speed up one run of the density peak algorithm using index, since multiple runs may be needed before getting a satisfactory clustering result.

\item \textit{List-based index.} We propose two list-based indices namely \textit{List Index} and \textit{CH Index} along with efficient query algorithms. An approximate solution is also suggested for memory-constrained systems to reduce the space cost. 

\item \textit{Tree-based index.} We revisit the popular tree-based indices having lower space cost and present efficient algorithms based on effective pruning techniques using the Quadtree Index and R-tree Index.

\item \textit{Extensive experimental evaluation.} Finally, we conduct extensive experiments to evaluate the proposed and existing indices on six datasets. 

\end{itemize}

The rest of the paper is organized as follows. Section 2 revisits the DPC method. In Section 3, we introduce our proposed list-based indices and their respective query algorithms, followed by the tree-based indices in Section 4. A comprehensive evaluation of the proposed index structures is done in Section 5. Section 6 discusses the related work. Lastly, Section 7 concludes the paper.

\section{Preliminary} \label{sec:densitypeaks}
In this section, we first introduce the Density Peak clustering method and then discuss the idea of our index-based approaches.

\begin{figure}
\centering
\subfloat[Data Distribution]{{\includegraphics[width=1.7in,height=1.3in]{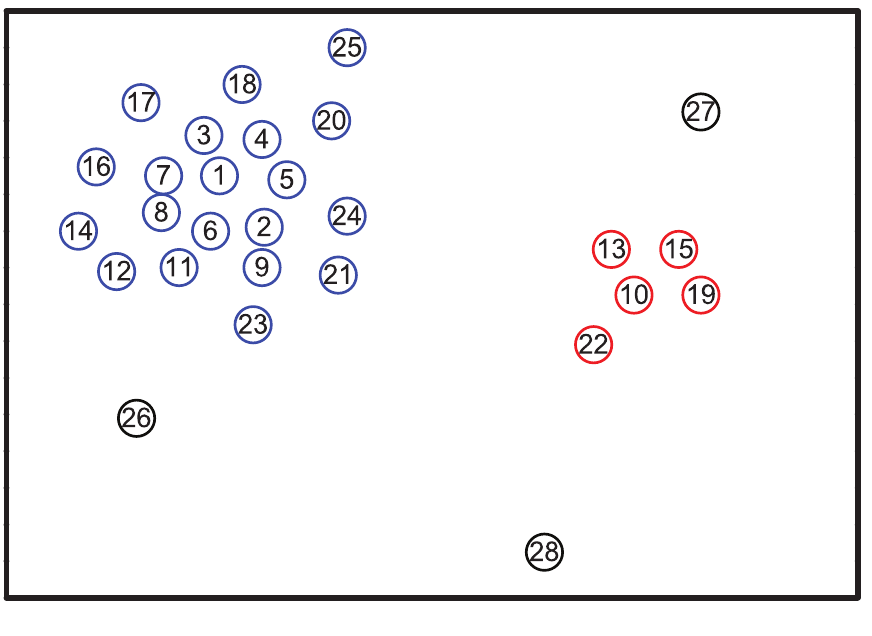}}
\label{fig:dpds}}
\hfill
\subfloat[Decision Graph]{{\includegraphics[width=1.7in,height=1.3in]{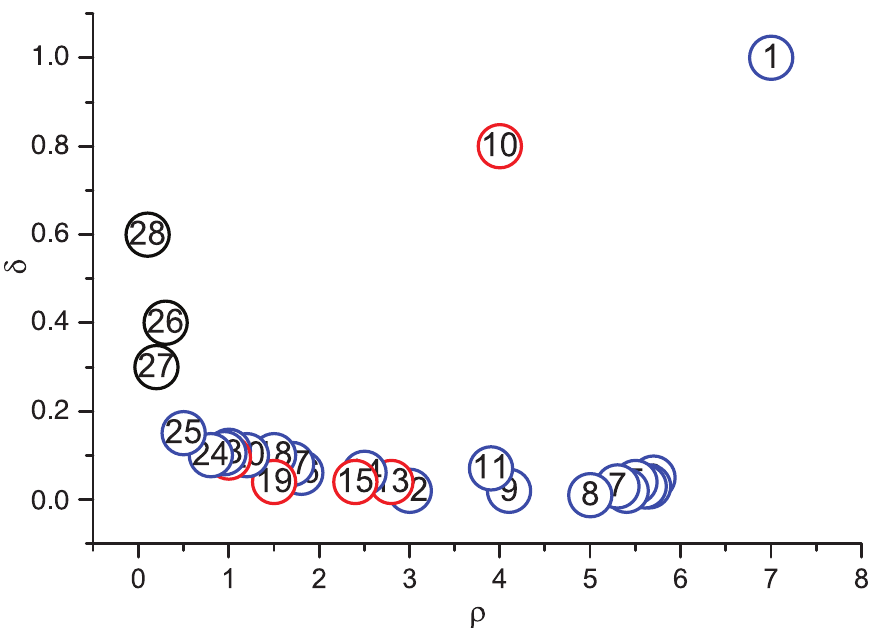}}
\label{fig:dpdg}}
\caption{Density Peak Clustering\protect\cite{rodriguez2014clustering}}
\label{fig:dpex}
\vspace{-5pt}
\end{figure}

\medskip
\noindent
\textbf{Density Peak Clustering (DPC):} %\label{dp}
DPC \cite{rodriguez2014clustering} is based on the observation that \textit{cluster centers} are characterized by 
(i) \textit{locally higher density}, i.e., a cluster center has a higher-density neighborhood than its neighboring objects, and 
(ii) \textit{relatively large separation}, i.e., cluster centers are at relatively large distances from other objects with higher local densities. On this basis, DPC distinguishes the cluster centers from the rest of the objects. Let $P$ be a set of objects to cluster, then the clustering procedure of DPC involves mainly the following four steps: 

\begin{enumerate}[leftmargin=*]
\setlength\itemsep{0.2em}
\item \textit{Compute local density $\rho_p$}. The local density $\rho_p$ of an object $p\in P$ is computed as

\begin{equation}{
\rho_p=\sum_{q\in P} \chi(dist(p,q) -d_c)
}\end{equation}
where $dist(p,q)$ is the distance between objects $p$ and $q$, $\chi(x)=1$ if $x<0$, otherwise $\chi(x)=0$, and $d_c$ is the threshold distance. %Basically, $\rho$ of an object \textit{p} is the number of objects that lie within distance $d_c$ from \textit{p}. 
We refer to the number of objects that lie within distance $d_c$ from $p$ as the (local) density $\rho_p$ of the object $p$.

\item  \textit{Compute dependent distance $\delta_p$}. The dependent distance $\delta_p$ is the minimum distance between object $p\in P$ and any other object $q \in P$ with density higher than $\rho_p$. It is computed as

\begin{equation}{
\delta_p= \min_{q \ne p \land \rho_q > \rho_p}\{dist(p,q)\}
}\end{equation}

We denote the corresponding higher density object (neighbor) by $\mu_p$. For the highest density object (global peak) $p$, its $\delta_p$ = $max_{q\in P}{\{dist(p,q)\}}$.

\item \textit{Finding cluster centers}.  
Then, DPC distinguishes cluster centers (peaks) from the set of objects based on their computed $\rho$ and $\delta$. An object with locally high density has its nearest neighbor of higher density relatively far and therefore has a large $\delta$. Based on this, cluster centers are recognized as objects with high $\rho$ and anomalously large $\delta$. For this, DPC employs a decision graph to determine the cluster centers.
Figure~\ref{fig:dpds} shows a data distribution of objects numbered according to the rank of their local densities. A decision graph of $\rho$ vs. $\delta$ is shown in Figure~\ref{fig:dpdg} which determines cluster centers (1,10) found on top right side of the graph with high $\rho$ and large $\delta$.  
The decision graph also helps to identify the outliers (26,27,28) which have small $\rho$ and large $\delta$ on the left side of the graph. 

\item \textit{Clustering.}
After the cluster centers have been determined, the rest of the objects are then assigned to the clusters containing their nearest neighbors of higher density.
\end{enumerate}

The time complexity of the original DPC algorithm \cite{rodriguez2014clustering} is $\Theta(n^2)$ dominated by the computation of pair-wise distance of objects. 
Then, computing $\rho$ and $\delta$ for all objects takes $\mathcal{O}(n^2)$ time. The third step, finding cluster centers, requires manual input from a user, after which, object-to-cluster assignment in the fourth step is done in $\mathcal{O}(n)$ time. 
Thus, the first two steps are most expensive. As DPC is sensitive to $d_c$, the DPC algorithm may need to be run multiple times. While the pair-wise distances can be reused after firstly computed, it would be preferred to make computing $\rho$ and $\delta$ more efficient, given that different $d_c$ values may be tried. Our index-based approaches are designed to support the computation of $\rho$ and $\delta$ efficiently.

%As DPC algorithm is sensitive to $d_c$, it entails that solutions be found where, for any $d_c$, redundant computations are not made. Additionally, metric queries should be significantly fast to produce the decision graph from which users can recognize the cluster centers and eventually obtain the clusters. Based on this, we introduce our index-based approach for DPC.

\medskip
\noindent
\textbf{Index-based Approach:} 
Our approach for DPC consists of applying an index and efficiently computing $\rho$ and $\delta$ (the first and second steps of the original algorithm) correctly based on the designed index. As the third and fourth steps are inexpensive and require user input, they can be used as they are in the original algorithm \cite{rodriguez2014clustering}. Based on this, we propose two list-based indices, namely List Index and Cumulative Histogram Index, along with efficient algorithms to compute $\rho$ and $\delta$ for each object. We also enable the popular tree-based indices Quadtree and R-tree for handling large datasets by developing enhanced algorithms based on effective pruning techniques for computing the two quantities. Once these quantities are obtained, 
cluster centers are determined using the decision graph, and finally, the object-to-cluster assignment is done by following the third and fourth steps of the original algorithm. 

The different indices are elucidated in the following sections. The notations used in this paper are listed in Table~\ref{tb:ntt} for reference.

\begin{table}
\centering
\renewcommand{\arraystretch}{1.1}
\caption{\\Summary of Notations}\label{tb:ntt}
%\vspace{-5pt}
\resizebox{0.90\columnwidth}{!}{%
\begin{tabular}{p{1.5cm} p{5.0cm}}
\hline
\textbf{Notation} & \textbf{Description}\\
\hline 
$d_c$ & cut off distance \\
$n$ & total number of data objects \\
$\rho_p$ & local density of an object $p$ \\
$\delta_p$ & dependent  distance of an object $p$\\
$\mu_p$ & higher density neighbor of an object $p$\\
$dist(p,q)$ & distance between objects $p$ and $q$\\
N-List & list of objects \\
RN-List & reduced list of objects \\
$\tau$ & neighbor threshold for RN-List \\
$w$ & width of histogram bins \\
$nc$ & number of objects inside tree node \\
$maxrho$ & maximum $\rho$ of objects inside tree node \\
$d_{min}$ & minimum distance to tree node \\
$d_{max}$ & maximum distance to tree node \\
\hline
\end{tabular}
}

\vspace{-8pt}
\end{table}

\section{List-based Index Structures}
The general idea behind list-based index is that the computation of $\rho$ and $\delta$ for an object requires neighboring objects to be retrieved and matched. However, in this process the query visits a large number of unnecessary objects.
Knowing the object's neighborhood beforehand can help to reduce the irrelevant comparisons significantly. List Index is based on the above idea and utilizes \textit{ordering} technique to maintain neighbors of each object in the order of their proximities. Using the List Index, efficient query algorithms for computing $\rho$ and $\delta$ are developed which find DPC clusters in $\mathcal{O}(n\log{}n)$ expected time. 
We also propose an advanced CH Index which utilizes \textit{aggregation} technique for enhancing the query processing of List Index. The CH Index integrates the merits of both cumulative histograms and list into its structure and achieves DPC clustering in just $\mathcal{O}(n)$ expected time. An overview of List and CH Index along with their construction and query algorithms will be explained next.

\subsection{List Index}\label{sec:list}
The List Index maintains a list known as \textit{Neighbor List (N-List)}
for each object, such that for an object \textit{p}, its N-List(p) stores other objects in non-decreasing order of their distances to $p$. This is useful, since, given a $d_c$, the portion of list containing $distance \geq d_c$ is irrelevant for the computation of $\rho_p$. Then, $\rho_p$ can be simply determined by just finding the location of the farthest object with $distance < d_c$ in N-List(p) for which a binary search is efficient.
The computation of $\delta$ is based on the observation that for peaks and outliers, $\delta$ is relatively large. This means that for a non-peak object $p$, its $\delta_p$ is small, i.e., its $\mu_p$ lies very close to it. Using N-List(p), which stores the nearest neighbor objects in the starting locations, $\delta_p$ can be determined by visiting only few starting objects of N-List(p). We explain the above computations for $\rho$ and $\delta$ using the following example.

\begin{example}
Figure~\ref{liststr} shows the N-List for objects {10, 13, 15, 19, 22} of the distribution in Figure~\ref{fig:dpds} containing other objects in non-decreasing order of their distances. Suppose $d_c=0.25$,
to find $\rho_{10}$ for object 10, the number of objects in N-List(10) up to the farthest object with $dist<0.25$ is required. Thus, we only need to find the location of that farthest object in N-List(10) which can be efficiently done by a binary search. Therefore, for object 10, the farthest object is 22 and its corresponding location gives the required $\rho_{10}=4$. Similarly, we obtain $\rho_{13}=3$, $\rho_{22}=1$, $\rho_{15}=3$ and $\rho_{19}=3$.
To compute $\delta$ for object 13, we find the first object in its N-List of higher density as object 10 (suppose a smaller object ID represents a higher local density). Thus, $\mu_{13}=10$ and the corresponding distance giving $\delta_{13}=0.12$ can be obtained in just one search. Similarly, for object 15, 19 and 22, $\delta$ can be obtained in 1-2 search operations. 
For object 10, a relatively large number of search operations are needed as there is no near object with higher $\rho$. Therefore, it has a large $\delta$.
\end{example}

\begin{figure}
\centering
\subfloat[List Index]{{\includegraphics[width=2.0in,height=1.3in]{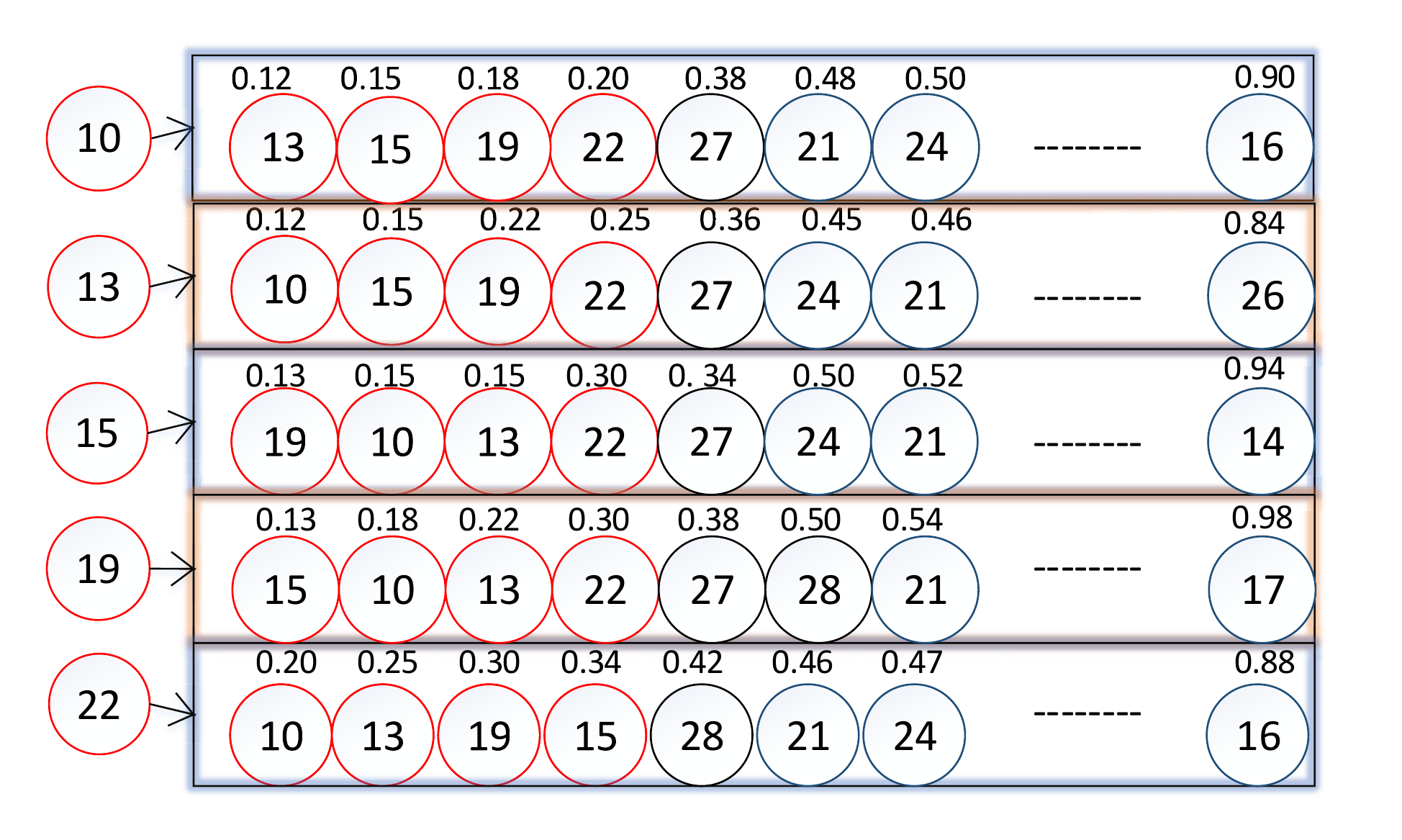}}
\label{liststr}}
\hspace{1pt}
\subfloat[Theorem~\ref{listproof}]{{\includegraphics[width=1.3in]{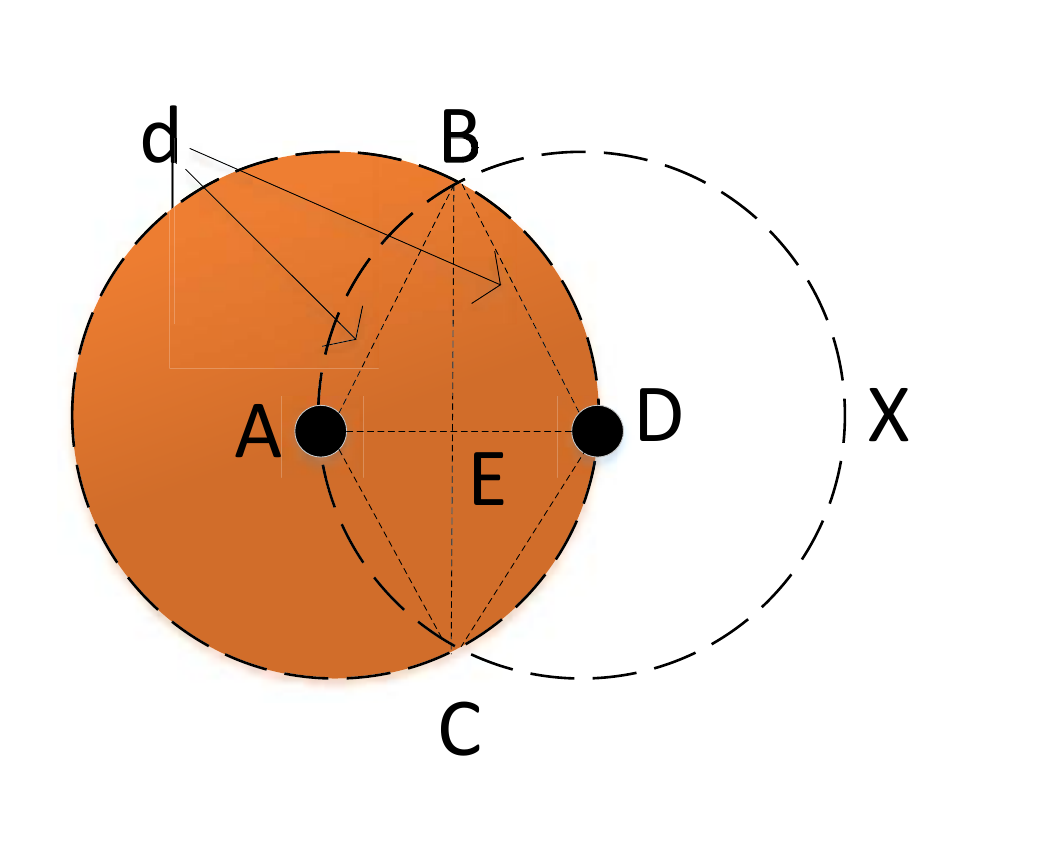}}
\label{fig:listproof}}
\caption{(a)List Index (b) For Theorem 1}\label{liststr}
\vspace{-8pt}
\end{figure}

\subsubsection{Construction}
The construction of List Index is shown in Algorithm~\ref{list1}. Line 1 initializes a List Index which stores the $\textit{list}$ of distances for each object. The algorithm first picks an object \textit{p}, computes the distances to all other objects and stores the objects (along with their distances $dist(p,q)$) in a temporary \textit{list} as shown in lines 3-6. This list is then sorted in non-decreasing order (line 6) and finally stored in List Index (line 8). This is repeated for all the objects and finally List Index is returned containing an N-List for each object. 

The time complexity of Algorithm~\ref{list1} is $\mathcal{O}(n^2 \log{} n)$ which includes the pair-wise distance computation of objects and sorting of each N-List.

\subsubsection{Query Algorithm}
The general idea of computing $\rho$ and $\delta$ using List Index has been discussed earlier.
The pseudo-code for both $\rho$ (lines 2-6) and $\delta$ (lines 7-13) queries is given in Algorithm ~\ref{alg:list2}. 
The algorithm initializes $\rho$-set, $\delta$-set and $\mu$-set in line 1 which store the set of $\rho_p$, $\delta_p$ and $\mu_p$ values for each object $p$. To compute $\rho_p$, it employs the efficient binary search technique over N-List(p), shown in line 5, to find
the location of object which is its corresponding $\rho$. 
The computed $\rho$ is stored in $\rho$-set in line 6. 

The query for computing $\delta$ is given in lines 7-13 of the algorithm. It iteratively picks an object $p$ and performs a sequential search over its N-List(p), each time checking if the current object $i$ has higher density 
shown in line 8-9. The search terminates (line 12) when the first object \textit{i} satisfying the condition is met, which is the required $\mu$, and its distance $dist(p,i)$ is the corresponding $\delta$.  
The obtained values are stored in their corresponding sets in line 13 after which the algorithm terminates, returning the final $\rho$-set, $\delta$-set, $\mu$-set. 

\begin{algorithm}[!t] 
\SetAlgoNoLine 
\relsize{-1}
\caption{List Index: Construction} \label{list1}
\KwIn{Set of objects of P}
\KwOut{List Index}
List Index $\gets$ Initialize list of list\;
$list$ $\gets$ Initialize a temporary list\;
\ForEach{object p $\in$ P}
{
\ForEach{object q $\in$ P}
{
$list(p) \gets q, dist(p,q)$\; 
}
N-List(p) $\gets$ sort $list(p)$ in non-decreasing  of $dist(p,q)$\;
Add N-List(p) into List Index\;
%\textit{List Index} $\gets$ N-List(p)\;
}
return List Index;
\end{algorithm}
\normalsize

\begin{algorithm}[!t] 
\SetAlgoNoLine
\relsize{-1}
\caption{List Index: Query}\label{alg:list2}
\KwIn{List Index, parameter $d_c$}
\KwOut{$\rho$-set, $\delta$-set, $\mu$-set}
Initialize $\rho$-set, $\delta$-set, $\mu$-set\;
\tcp{Computing $\rho$ }
\ForEach{object p $\in$ D}{
first $\gets$ 0; \tcp{Initialize first to 0}
last $\gets$ N-List(p).length()-1; \tcp{Initialize last to location of last element of list}
$\rho_p$ $\gets$ BinarySearch(first, last, N-List(p), $d_c$)\;
$\rho$-set $\gets \rho_p$\;
}
\tcp{Computing $\delta$}
\ForEach{object p $\in$ D}{
\ForEach{object q $\in$ N-List(p))}{ 
\uIf{($\rho_q > \rho_p$)} {
$\delta_p \gets dist(p,q)$\;
$\mu_p \gets q$ \;
$break$\;
}
}
$\delta$-set $\gets \delta_p$, $\mu$-set $\gets \mu_p$\;
}
return $\rho$-set, $\delta$-set, $\mu$-set;
\end{algorithm}
\normalsize

\begin{theorem} \label{listproof}
\itshape The expected time complexity of the Algorithm~\ref{alg:list2} is $\mathcal{O}(n \log{} n)$.
\end{theorem}
\vspace{-2pt}
\begin{IEEEproof}
\hangindent=2em
Let $P$ be a set of $n$ objects and $p, q \in P$. The time complexity for computing $\rho$ in lines 2-6 is $\mathcal{O}(n \log{} n)$. This is because, for each object $p \in P$, binary search performs $\mathcal{O}(\log{} n)$ comparisons. 
\hangindent=2em 
The expected time complexity for computing $\delta$ in lines 7-12 is $\mathcal{O}(1)$. To prove this, we assume that, for each cluster, the density is maximal at the peak object and the density of object decreases while moving away from the peak.
To compute $\delta$ for $p$, the query probes each object from near to far in its N-List until it finds the first object \textit{q} with higher density. Suppose query object $p$ and object $q$ are located at points D and A as shown in Figure~\ref{fig:listproof}. For the non-peak object $p$, the area defined by circular region centered at D and radius $dist(D,A)$ is the total area that the query needs to explore to find $\delta$ defined as \textit{area(near)+area(far)}. The area(near) is the dense area in between $p$ and $q$, while area(far) is the remaining area which is sparse. The ratio of area(far)/area(near) is no more than a constant $f=\frac{\pi/3 + \sqrt3/2}{2\pi/3 - \sqrt3/2}$. This means, before one higher density object is found in area(near), the number of lower density objects checked in area(far) is in a constant factor. Therefore, the total number of objects probed for an object is expected to be a constant number.  
For peak objects, since the higher density neighbors are not near, in the worst case, the number of probes is $n$. Assuming the number of peaks as constant $c$, the total number of probes is bounded by $c n$. Thus, finding $\delta$ for all the objects requires total $f(n-c)+cn$ probes which cost $\mathcal{O}(n)$ expected time. Hence, overall expected time complexity of Algorithm~\ref{alg:list2} is $\mathcal{O}(n \log{} n)$ + $\mathcal{O}(n)$ = $\mathcal{O}(n \log{} n)$. 
\end{IEEEproof}

\begin{figure}
\center
\includegraphics[width=2.7in]{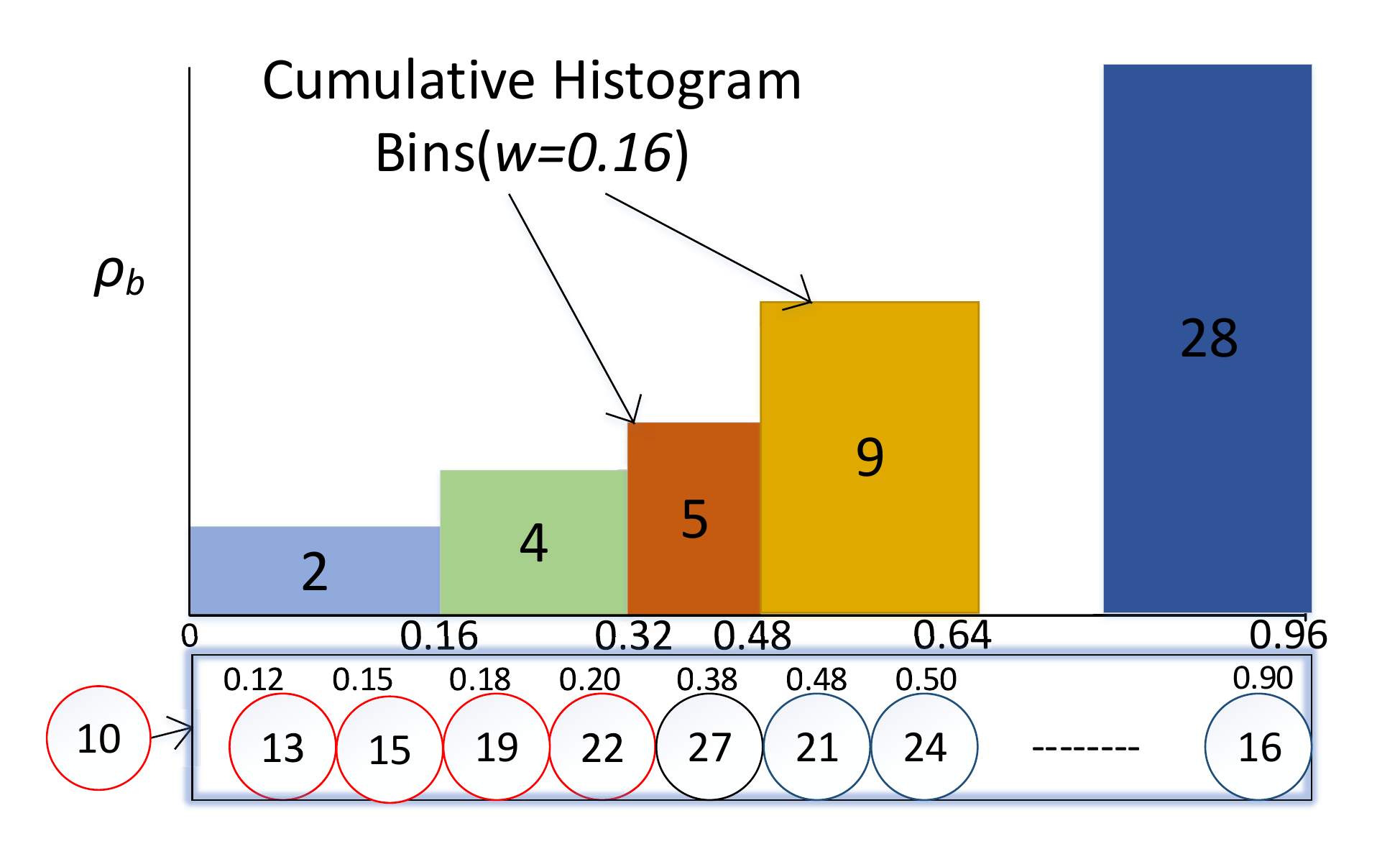}
\caption{Cumulative Histogram for Object 10 in Figure~\ref{fig:dpex}}
\label{fig:histgen}
\vspace{-10pt}
\end{figure}

\subsection{Cumulative Histogram}
List Index can speed up the DPC queries, however, the computation of $\rho$ is affected by the size of N-List. With increase in dataset size, the size of an N-List grows larger. Consequently, more comparisons are performed, thereby making the query slow. 
To this end, we introduce a Cumulative Histogram (CH) Index for DPC which takes $\mathcal{O}(1)$ time to compute $\rho$ for an object. CH Index merges the merits of both \textit{List Index} and \textit{cumulative histogram} which helps to tremendously reduce the search space of List Index. The basic idea is to provide smaller search space by dividing N-List of an object into small subsets such that the number of comparisons during the computation of $\rho$ is reduced.

CH Index contains a cumulative histogram for each object consisting of several bins where each bin represents a disjoint range of distance w.r.t. $p$. For an object $p$, the first bin indicates the number of objects ($n_b$) in N-List within distance $w$ from $p$, where $w$ is the \textit{bin width} and is user defined. Similarly, the second bin indicates distance $2w$ from $p$, and so on, until the bins cover all the objects of N-List. Distances $w$, $2w$, $3w$,... are the \textit{upper limit} of first, second, third bin respectively and so on, while $w$ is the \textit{bin width} of each bin. 
Since the number of objects within distance $kw$ from $p$ in N-List(p) is equal to the location of the last object $q$ with $dist(p,q)<kw$, each bin stores the location of the last object as \textit{bin density} ($n_b$). To compute $\rho$ for an object $p$, the query needs to locate the bin (say $targetBin$) containing the farthest object $q$ with $dist(p,q)<d_c$, and perform a search in the farthest bin. 

A smaller \textit{bin width} $w$ represents a smaller section of N-List and results in faster query time (at the cost of additional space as there will be more bins) and vice-versa. Thus, for a particular dataset, selecting an appropriate $w$ is of paramount importance to both the running time and space cost, and depends on the choice of user.

\begin{example}
Figure~\ref{fig:histgen} represents a cumulative histogram for object 10 along with its N-List. 
It is constructed by counting the number of objects in N-List with distance less than \textit{upper limit} of each bin (shown at the abscissa, for 0.16, 0.32, 0.48, 0.64,.., 0.96) which is stored at each bin as $n_b$. The last bin contains the total number of objects in N-List.
Given $d_c=0.25$, firstly the \textit{targetBin T} is determined by calculating $\floor{d_c/w} = 1$. In the figure, the second (green) bin is the \textit{targetBin} as $d_c$ lies within [0.16, 0.32). Here, $n_T=4$ and $n_{T-1}=2$ (previous bin) determines the section of N-List which needs to be explored to find $\rho$ of object 10. A search is performed on this section of N-List containing only 2 objects (19, 22) which returns the location of object 22 as the corresponding $\rho$. It must be noted that List Index has to search over 28 objects while CH Index just needs 2 objects in this case. 
\end{example}

\begin{algorithm}[!t]
\SetAlgoNoLine
\relsize{-1}
\caption{Histogram Construction} \label{alg:hist}
\KwIn{List Index, width(w)}
\KwOut{CH Index}
CH Index $\gets$ Intialize list of lists\;
\ForEach{object p $\in$ D}{
upper limit $\gets$ w; \tcp{Initialize first bin's upper limit}
i $\gets$ 0\;
c\_histogram $\gets$ initialize list for storing bin density\; 
\While{i $<$ N-List[p].size()}{
dist(p,i) $\gets$ N-List[p][i]\;
\uIf{$dist(p,i) < upper\ limit$}{
$i \gets i+1$\;
}
\Else{
$c\_histogram.insert(i)$\;
\textit{upper limit} $\gets$ \textit{upper limit}+$w$\;
}

}
$c\_histogram.insert(i)$\;
CH Index[p] $\gets$ c\_histogram\;
}
return CH Index;
\end{algorithm}
\normalsize

\subsubsection{Construction}
The construction of CH Index is given in Algorithm ~\ref{alg:hist}. The algorithm uses the List Index to construct cumulative histograms for each object in lines 2-14.
For each object $p$, it initializes \textit{upper limit} of the first bin with $w$, \textit{i} to iterate over N-List, and a list \textit{c\_histogram} to store $n_b$ for each bin as shown in lines 3-5. 
To find $n_b$ of the first bin, the algorithm iterates over N-List(p) and examines for object $i$ whether $dist(p,i)$ $<$ \textit{upper limit} as shown in line 7-8. If it falls within the first bin, it increments $i$ in line 9 and moves to next object of N-List. This is repeated until the first object at location $\textit{i}$ with $dist(p,i)$ $>$ \textit{upper limit} is met. The location $\textit{i}$, which at this instance gives the number of objects within distance $w$ from $p$, is stored as $n_b$ of the first bin in c\_histogram as shown in line 11. Line 12 increments \textit{upper limit} by $w$, which indicates the \textit{upper limit} of the second bin. The algorithm again starts to examine the objects from the last location \textit{i} and repeats the same procedure to store the $n_b$ of the second bin. 

In this way, the loop executes until all the objects of N-List(p) have been visited. Finally, line 13 stores the the $n_b$ of the last bin. The list containing the $n_b$ values (i.e., c\_histogram) of object $p$ is stored in CH Index in line 14. The algorithm then moves to another object and iterates similarly over its N-List and constructs the bins. When bins for all the objects have been constructed, the algorithm terminates returning CH Index (line 14).
The total time complexity to construct CH Index based on Algorithm~\ref{list1} and ~\ref{alg:hist} is $\mathcal{O}(n^2 \log{} n)$.

\subsubsection{Query Algorithm}

In this section, we describe the algorithm to compute $\rho$ using CH Index.
For an object $p$, the algorithm finds the $targetBin$ to perform a search on the corresponding section of N-List(p), which finds the location of the farthest object $q$ with $dist(p,q)<d_c$ as the required $\rho$. 

The pseudo-code for computing $\rho$ is shown in Algorithm~\ref{alg:hist1}. 
For each object $p$, the \textit{targetBin} is computed as shown in line 1-2. 
Next, the algorithm computes $\rho$ for $p$ in lines 3-17. If $d_c$ is equal to \textit{upper limit} of $targetBin$, all objects up to that bin are within distance $d_c$. Thus, $n_b$ of $targetBin$ is directly assigned to $\rho_p$ as shown in line 5-6.
Otherwise, in lines 8-14, the algorithm performs a search on the section of N-List corresponding to $targetBin$. Since each bin contains the location of object in N-List, first and last are retrieved from the previous and current bin respectively as in lines 12-13. %Obviously if targetBin = 0, first = 0. 
If the $targetBin$ is greater than the total number of bins, all the objects are within $d_c$ and $n_b$ of the last bin is assigned to $\rho$ as shown in line 17. The $\rho_p$ obtained is stored in $\rho$-set in line 18. The same procedure is repeated to compute $\rho$ for all the objects which are stored in $\rho$-set after which the algorithm terminates.

\begin{theorem}
\itshape The time complexity of Algorithm~\ref{alg:hist1} is $\mathcal{O}(n)$.
\end{theorem}
\vspace{-2pt}
\begin{IEEEproof}
\hangindent=2em
The algorithm takes $\mathcal{O}(1)$ time to locate the targetBin in lines 1-2. Then, if say, there are $b$ objects of N-List to be explored between $first$ and $last$ locations, line 14 takes $\mathcal{O}(b)$ time to compute $\rho_p$. A careful chosen $w$ will make $b$ near-constant, and thus the time complexity can be regarded as $\mathcal{O}(1)$. Therefore, the overall time complexity of Algorithm~\ref{alg:hist1} is $\mathcal{O}(n)$.  
\end{IEEEproof}

\begin{theorem}
\itshape Given any $d_c$, Algorithm~\ref{alg:list2} and Algorithm~\ref{alg:hist1} compute correctly the set of all clusters in D.
\end{theorem}
\vspace{-2pt}
\begin{IEEEproof} 
\hangindent=2em
Let's assume that our algorithm returned a set of clusters. 
Since, the correctness depends upon the set of cluster centers obtained, which further depends on the computed $\rho$ and $\delta$ values,
the algorithm should obtain correct $\rho$ and $\delta$ for any given $d_c$.
As an object's N-List stores objects in non-decreasing order of their distances, a search through its N-List finds the location of another object with a distance just smaller than $d_c$. As the objects are in sorted order, the list guarantees that all the objects stored before $d_c$ lie within the range $d_c$ and thus local density $\rho$ of that object is obtained. Thus, Algorithm ~\ref{alg:list2} successfully obtains correct values of $\rho$ for all objects. Similarly, cumulative histogram provides a small subsection of list containing $d_c$. A cheap search is applied as above which returns the required location. Therefore, Algorithm~\ref{alg:hist1} also obtains correct $\rho$ for all objects.
To compute $\delta$, the algorithm performs a linear search which returns the first object with higher $\rho$ in a list. As $\rho$ has been computed correctly, performing a sequential search guarantees to find correct $\mu$. Based on these values, correct cluster centers are determined and hence clustering results are correct.
\end{IEEEproof}

\begin{algorithm}[!t] 
\SetAlgoNoLine
\relsize{-1}
\caption{$\rho-Query$ using CH Index} \label{alg:hist1}
\KwIn{CH Index, $d_c>0$}
\KwOut{$\rho$-set : Set of $\rho$ for all Objects}

$bin \gets d_c / binsize$\; 
$targetBin \gets floor(bin)$\;
\ForEach{object p $\in$ D}{
\uIf{$targetBin < histogram[p].size()$}
{
\uIf{bin = targetBin} 
{\tcp{targetbin equals bin's upper limit}

$\rho_p$ $\gets$ $c\_histogram[p][targetbin-1]$\; 
}
\Else
{
\uIf{targetBin = 0}{
first $\gets$ 0\;
last $\gets$ $c\_histogram[p][targetBin]-1$\;
}
\Else{
first $\gets$ $c\_histogram[p][targetBin-1]$\;
last $\gets$ $c\_histogram[p][targetBin]-1$\; 
}
$\rho_p$ $\gets$ Search(first, last, N-List(p))\;

}
}
\Else{
$\rho_p$ $\gets$ c\_histogram[p][last]\;

}
$\rho$-set[p] $\gets \rho_p$\;
}
return $\rho$-set; 
\end{algorithm}
\normalsize

\subsection{Approximate Solution}
We present an approximate solution to adapt the above indices to work with low memory systems while still providing fast run time. The approximate solution significantly reduces memory requirement of the indices at the cost of slight inaccuracies in the clustering results. 

The concept driving this is that quantities $\rho$ and $\delta$ mainly require visiting neighboring objects. An N-List contains a large number of near and far objects where the far objects with  distance $>d_c$ do not contribute to $\rho$. Similarly, for most objects, their $\mu$ objects are near and usually easy to be found in starting locations of N-List. This implies that far neighbors are not meaningful and storing them costs unnecessary space for most objects. Therefore, if the maximum relevant neighborhood regions of the objects are known, only the objects of these regions need to be stored.

Building on the above idea, we introduce \textit{Reduced Neighbor List (RN-List)} with a $neighbor\ threshold$ parameter $\tau$. Using the stored N-List, RN-List of an object retrieves and stores only those objects with distance $<\tau$, where $\tau$ defines the radius of maximum relevant neighborhood region, and the queries are executed using RN-List instead of N-List. 
Usually $\tau$ should be set to a large value greater than any possible value of $d_c$ to be tested by user. This helps to obtain correct $\rho$ for any $d_c$. Given correct $\rho$ and large $\tau$, correct $\delta$ can be obtained for non-peak objects as they have smaller $\delta$. For peak objects with $\delta>\tau$, their $\delta$ is set to a large value. This simple setting helps to find cluster centers with high $\rho$ and anomalously large $\delta$. Note that, a very small $\tau$ may change most objects to cluster centers/outliers. Thus, it is important that a sufficiently large $\tau$ is selected to obtain near accurate results. Interestingly, it was observed that for some datasets, less than 1\% of the total number of objects were probed using List Index. 

However, storage cost may still be high and List Index may not ensure quality for very large datasets, as the size of RN-List diminishes. In addition, it is also difficult to provide approximation ratios due to different data distributions of different datasets. This motivates the use of tree-based index, as introduced in the subsequent section.

\section{Tree-based Index Structures}
List-based indices significantly accelerate the query time of DPC, but require storing neighbors for each object which has high space cost. For low memory systems, it may not be possible to store list-based index for large datasets. Therefore, users cannot exploit their advantages to find DPC clusters efficiently. It would be beneficial if some indices could support efficient queries along with low space requirement. Moreover, the preprocessing cost of list-based indices is also high especially for large datasets. Although such indices are constructed only once, this high cost may be undesirable where a user wants fast construction of index before obtaining clustering results efficiently.  

To this end, we study tree-based indices which have low memory requirements 
and efficient neighborhood search queries.
As discussed earlier, we concentrate on popular Quadtree Index and R-tree Index. These indices are well suited to the problem of DPC as they can easily eliminate large irrelevant regions from consideration. 
This improves the execution time of queries like $\rho$ and $\delta$, intending to seek information about the neighborhood. Thus, DPC clusters can be obtained using tree-based indices. 

The popular range search and nearest neighbor search (\textit{NNS}) are efficient algorithms for their purposes. Nevertheless, they cannot be directly adapted for the computation of DPC quantities, particularly $\delta$. 
%The range search can answer the $\rho$ query, however, it requires substantial number of search operations, even for a moderately large $d_c$. 
The computation of $\delta$ is different from \textit{NN} search, where the nearest object is retrieved. We are interested in finding the higher density nearest neighbor which may be different from the normal nearest neighbor. This is true especially for peak objects for which the general query will end up searching a large search space. These queries are slow in finding DPC clusters, which motivates developing efficient algorithms for the same. In the following subsections, we first discuss the Quadtree Index, and present efficient query algorithms to compute $\rho$ and $\delta$ based on two pruning techniques. Then, we discuss the R-tree Index.

\subsection{Quadtree}
A Quadtree involves a hierarchical decomposition of the space in a regular manner.
The root of the tree is congruous to the whole space in consideration, and the levels below root depict a refinement of space. In a quadtree, each non-leaf node has four children and the leaf node contains the actual data objects. These children are obtained by recursively subdividing the space into four regions. The division of node occurs when the number of objects increases beyond its maximum capacity. However, as the structure of Quadtree depends on the distribution of objects, in the worst case, its height may become linear with the number of objects. 
The pruning techniques and query algorithms for computing the DPC quantities using Quadtree are discussed below.

\subsubsection{Pruning Techniques} 
  The first observation helps to improve the computation cost of $\rho$ by pruning several tree nodes. Let $P$ be a set of objects such that $p,q \in P$, function $d_{min}(p,node)$  determine the minimum distance of the object to the region defined by the node and $d_{max}(p,node)$ determine the maximum distance to that region, then 
 
 \begin{observation}
\itshape Given an object $p \in P$, a query, defined by circular region Q centered at p and radius $d_c$,  determine whether a tree node representing region R is
\textit{fully contained,} if $R\subset Q$;   %completely overlaps the node,
\textit{discarded,} if $R\cap Q=\emptyset$; or
\textit{explored} otherwise.
\label{obs1}
\end{observation}

\medskip \noindent
This is straightforward because if $R$ is entirely within $Q$ i.e., $R\subset Q$, all objects in $R$ are guaranteed to be in $Q$ and can be directly added to $\rho$. This can be verified by checking if $d_{max}(p,node)<d_c$. If $R$ is completely outside $Q$, i.e., $R\cap Q=\emptyset$, no object of $R$ lies in $Q$. This can be found by checking if $d_{min}(p,node)\geq d_c$. In both the cases, there is no need to explore these nodes. However, if $Q$ intersects the node, the number of objects in $R$ also within $Q$ needs to be determined, and thus child tree nodes (sub-regions) need to be explored. Next, we present two lemmas which help prune the irrelevant nodes during the computation of $\delta$.
\begin{lemma}\label{lemma1}
\itshape (Density Pruning) Given an object $p \in P$ with local density $\rho_p$, its $\mu$ does not lie in a node of Quadtree which satisfy $maxrho (node) < \rho_p$ where $maxrho(node)$ is the maximum local density of an object inside that node.
\end{lemma}

Clearly, for an object $p$, its $\mu$ cannot be found in nodes with no object $q$ such that $\rho_q > \rho_p$. Therefore, such nodes are \textit{density pruned}. This is highly effectual in case of peak objects for which many unnecessary nodes as well as objects of lower densities can be pruned.

\begin{lemma}\label{lemma2}
\itshape (Distance Pruning) Given an object $p \in P$, its $\mu$ does not lie in a node of Quadtree which satisfies $d_{min} > candidate(\delta)$,
where $candidate(\delta)$ is the best $\delta$ obtained before visiting this node.
\end{lemma}

Once a candidate $\delta$ has been obtained, the next nodes that have $d_{min}(p,node) > candidate(\delta)$ are guaranteed to not provide better $\delta$. Such nodes can be $\textit{distance pruned}$. Based on these pruning techniques, efficient query algorithms are designed to compute the two quantities. To summarize, we store $nc$ (denoting the number of objects contained in the region of a tree node) and $maxrho$ at the nodes, which helps in computing $\rho$ and $\delta$ respectively.

\subsubsection{Construction}
The construction of Quadtree is simple and follows hierarchical decomposition of the space into four children at each level as discussed earlier. For insertions, the query performs comparisons at each level of quadtree and decides the relevant subtree to be followed. The object is inserted finally at the leaf node of the corresponding subtree.  
The $nc$ of each tree node can be computed when building the tree. The count is increased by one, when an object falls into the region.

The time complexity for the construction of Quadtree depends upon the resulting Quadtree. For random insertion, the average time spent is $\mathcal{O}(n \log_4 n)$. However, the worst-case where objects are inserted only in the deepest node can cost $\mathcal{O}(n^2)$ time. %Maintaining $\textit{nc}$ at each node requires traversing the complete tree once which is proportional to number of nodes.

\begin{algorithm}[t]
\SetAlgoNoLine
\relsize{-1}
\caption{Query Algorithm for $\rho$} \label{alg:qrho}
\KwIn{Object $p$, distance $d_c$, current node}
\KwOut{ $\rho$}
\SetKwProg{Fn}{Function}{:}{end}
\Fn{computeRho(p, $d_c$, node)}{
%compute dmin(p,e), dmax(p,e)\;
\uIf{node is leaf}{
\ForEach{object q $\in$ node}{
\uIf{$dist(q,p) < d_c$}{
count = count+1;
}
}
return count;
}
\uElse{
\uIf{$d_{min}(p,node) >= d_c$}{ 
return 0; \tcp{discard}
}
\uElseIf{$d_{max}(p,node) < d_c$}{ 
return nc; \tcp{fully contained}
}
\uElse{ \tcp{intersects}
\ForEach{children c of node}{
$\rho$ $+=$ $computeRho(p, d_c, c)$
\;

}
return $\rho$\;
}
}
}
\end{algorithm}
\normalsize

\subsubsection{Query Algorithm}
We first present a general idea of the query algorithms for computing the DPC quantities.
To compute $\rho$ of an object, the algorithm visits the root of Quadtree and performs checking of candidate nodes at each level by categorizing the tree node as fully contained, intersected or discarded based on Observation~\ref{obs1}. The \textit{nc} of a fully contained node is directly 
%A \textit{fully contained} node is pruned and its \textit{nc} is directly
added to $\rho$. Secondly, a discarded node is simply pruned. Only the intersected nodes at each level are explored. At the leaf nodes, the query counts the number of objects $q$ with $dist(p,q)<d_c$ and adds to $\rho$. The final $\rho$ obtained is the required answer.

To compute $\delta$ for an object, we employ the best first search heuristic which
visits the candidate node first that has a higher chance of finding $\mu$. This node apart from not pruned by Lemmas~\ref{lemma1},~\ref{lemma2} should also be the nearest one among all candidate nodes. This helps to find the first candidate $\delta$ quickly based on which 
the candidate nodes with $d_{min}(p,node)>\delta$ can be pruned. If a better $\delta$ is obtained in other unpruned candidate nodes, the candidate $\delta$ is updated. This is repeated until all candidate nodes have either been explored or pruned. The final candidate $\delta$ is the required answer.

Now we discuss these algorithms in detail. Below, $d_{min}$ and $d_{max}$ for an object $p$ are calculated by functions $minDistToNode(p,node)$ and $maxDistToNode(p,node)$ respectively.  

\medskip\noindent
\textbf{Algorithm for $\rho$.} The pseudo-code for computing $\rho$ of object $p$ is shown in Algorithm~\ref{alg:qrho} and uses function computeRho() with parameters $p$, $d_c$ and current node (initially root), as shown in line 1 which performs a depth first traversal over the nodes of Quadtree.
The query starts from root node and examines whether it is discarded, fully contained or intersected in lines 8, 10, 12 respectively. If the root node is fully contained, its \textit{nc} is returned as the required $\rho$ and the algorithm terminates. Since query object $p$ is within root i.e., $d_{min}(p,node)=0$, it is not discarded. The root node is explored only if it is intersected by the query and its children are visited by recursively calling the function computeRho() as shown in line 14. The current child node of root is again examined with the three conditions and is explored only if it is intersected by query. This continues until the leaf node is reached. If the current node is leaf, the query counts the number of objects with   $dist(p,q)<d_c$ as shown in lines 3-5. The count is returned to the calling function (line 14). In a similar manner, the algorithm explores all the intersected nodes at each level and terminates after returning the required $\rho$ (line 15). 

\medskip
\noindent
\textbf{Algorithm for $\delta$.} To compute $\delta$, firstly a simple post-order traversal algorithm is performed which maintains the pruning information based on Lemma~\ref{lemma1}. For the leaf nodes, the query finds the object with highest local density and stores its value at that node as \textit{maxrho}. For the internal node, it finds its \textit{maxrho}
by finding the maximum \textit{maxrho} of its children. Based on this information stored at each node, the algorithm to compute $\delta$ is explained.  

 \begin{algorithm}[t]
\SetAlgoNoLine
\relsize{-1}
\caption{Query Algorithm for $\delta$} \label{alg:qdelta}
\KwIn{Object $p$, Tree node}
\KwOut{$\delta_p, \mu_p$}
$stack <TreeNode> nodes$\;
$\delta_p \gets$ inf; $root.d_{min}$ $\gets$ 0\; 
nodes.push(root, root.$d_{min}$)\;
\While{nodes is not empty}{
$tp$ $\gets$ nodes.pop()\; 
\tcp{Distance Pruning}\
\uIf{$tp.d_{min} < \delta_p$}{ 
\uIf{node is Leaf}{
\ForEach{object q in node}{
\uIf{$\rho_q > \rho_p$} {
\uIf{$\delta_p > dist(p,q)$}{
$\delta_p \gets dist(p,q)$; $\mu_p \gets q$ \;
}
}
}
} %End If
\Else{
$temp.d_{min}$ $\gets$ inf; $temp.node$ $\gets$ null\;
\ForEach{child c of tp.node}{
\tcp{Density Pruning}\
\uIf{$c.maxrho > \rho_p$} {
\uIf{$temp.d_{min}$ $\leq$ $d_{min}(p,c)$}{
nodes.push($c$, $c.d_{min}$)\;
}
\Else{
\uIf{$temp.d_{min}$ != inf}{
nodes.push($temp.node$,$temp.d_{min}$);}
$temp.d_{min}$ $\gets$ c.$d_{min}$\;
$temp.node$ $\gets$ c\;
}
}
}
\uIf{temp.node != null}{
nodes.push($temp.node$, $temp.d_{min}$)\;
} 
} %End Else
}
}
return $\delta_p$, $\mu_p$\;

\end{algorithm}
\normalsize

The pseudo code to compute $\delta$ is shown in Algorithm~\ref{alg:qdelta}. The inputs are object $p$ and node of the tree (initially root). Line 1 initializes a stack of TreeNodes which contains
nodes along with their minimum distances $d_{min}$ from object $q$. Line 2 initializes $\delta_p$ to Infinity. We call $\delta_p$ a candidate. 
The algorithm first starts from root node, assigns zero to $d_{min}$, as the query object $q$ is within the region defined by root node, and then pushes it into stack. 
Line 3 pushes the root to the top of stack. The while loop (lines 4-24) terminates when the \textit{stack} is empty.
The top element of stack (now root) is popped out in line 5 and is examined for distance pruning i.e., if its $d_{min} < \delta_p$ in line 6. Since at this instance, $\delta_p$ is infinity, the node is not pruned and the algorithm proceeds to check if it is a leaf or non-leaf node in line 7.
As node is not leaf, lines 13-24 are executed. In line 13, a temporary node \textit{temp} is initialized with node as $null$ and $d_{min}$ as infinity. The purpose of this \textit{temp} node is to find the best node from the candidate nodes i.e., root's children as shown in lines 14-24. Density pruning is performed to only check those nodes that have higher \textit{maxrho} as shown in line 15. Then
$d_{min}$ is computed for all such nodes and the algorithm pushes the node with the smallest $d_{min}$ to the top and the rest after it without any preference (lines 16-24). Note that, here a priority queue can be used to replace the stack.

The best node is explored further whose children are again checked using density pruning, and a best node is selected. This is repeated until it reaches the leaf where each object is examined to find the one with higher $\rho$ shown in line 9. The distance to the first object with higher $\rho$ is recorded if it is better than the current candidate $\delta$. If so, query updates the candidate $\delta_p$ and $\mu_p$. 
 %Remember, at this instance, top element in \textit{stack} is one of the last candidate nodes which was filtered from density pruning step in line 15. 
 Then the next node of \textit{stack} is popped and examined to check whether its $d_{min}$ is smaller than candidate $\delta_p$ in line 6, i.e., has a chance of finding a better candidate $\delta_p$. If so, the algorithm performs the same procedure shown in lines 7-24. Otherwise, the algorithm moves to next element of \textit{stack} and examines it. In this way, the algorithm pops and examines all nodes of \textit{stack}. Once the \textit{stack} is empty, while loop is terminated. The final candidate $\delta_p$ and $\mu_p$ are the required results.
 
 \subsection{R-tree}
Although Quadtree is a simple and powerful index structure, its height is sensitive to the distribution of objects, and the tree can be unbalanced. This results in declining query performance for the neighborhood search queries. Even the pruning strategies may not be very helpful in such scenarios. This requires a need to explore other index structures with better compactness guarantee. One such structure is R-tree proposed by Guttman~\cite{guttman1984r}.

R-tree is an extension of B-tree for multidimensional objects. It is a balanced tree and is also based on hierarchical decomposition of data. In an R-tree, a node contains multiple entries of the form (rect, ptr). At each node, \textit{rect} is the \textit{minimum bounding rectangle (mbr)} of the enclosed objects in space and \textit{ptr} is the pointer to it. The root of the R-tree must contain at least 2 children unless it is a leaf. Thus, the height of an R-tree indexing $n$ objects is $\mathcal{O}(log{}_M n)$ where \textit{M} is the number of entries of a tree node.

However, the nodes of an R-tree may suffer from node \textit{overlap} and \textit{region coverage} by \textit{mbr}. Overlapping increases chances of unnecessary path search while a large region covered by an \textit{mbr} reduces the performance of pruning. To this end, several variants of R-tree
  ~\cite{beckmann1990r,leutenegger1997str,garcia1998greedy} 
were proposed to improve the query performance of R-tree by reducing overlap and region coverage. These variants basically differ in the method of constructing an R-tree but the traversal and search queries are the same as the original R-tree. 
Among the variants, the packing algorithm~\cite{leutenegger1997str,garcia1998greedy} often results in better structure with typically less overlap and better storage utilization as compared to other variants which results in improved query performances. 

\subsubsection{Construction}
We describe the basic idea of the construction using packing algorithm~\cite{leutenegger1997str} which is based on bulk-loading of objects.
The packing algorithm requires data to be preprocessed first before loading. Suppose, $n$ objects of data are to be inserted. The overall strategy is to recursively split the data space into small partitions, and eventually each partition is stored at a leaf node. These leaf nodes are further grouped together to form the non-leaf or internal nodes of the tree. Lastly, the grouping stops when the root node is created.

The partitioning strategy is explained as follows. If M is the maximum capacity of a tree node and L is the number of leaf nodes where $L = \ceil{n/M}$, the partitioning strategy for the first split is to sort the objects by the first dimension $x$ and divide them into $ \ceil{\sqrt{L}}$ partitions where each partition contains $M \times \ceil{\sqrt{L}}$ objects. Similarly, these partitions are recursively split by sorting the objects of each partition using the other dimension $y$ until each partition contains maximum \textit{M} objects (assuming there are two dimensions). 

\subsubsection{Query Algorithm}
Since both Quadtree and R-tree nodes represent regions of space bounding a set of objects, the pruning techniques are still valid for R-tree. The pruning techniques and query algorithms for DPC quantities can be easily adapted to the above constructed R-tree. Therefore, we omit the discussions of the respective algorithms in this section. The time complexities of computing $\rho$ and $\delta$ are summarized using the following lemma.

\begin{theorem}
\itshape The average time complexity of DPC algorithm using R-tree is \textbf{$\mathcal{O}(n \log_M{} n)$}.
\end{theorem}
\vspace{-2pt}
\begin{IEEEproof}
\hangindent=2em
A range query with R-tree takes $\mathcal{O}(\log_M{} n)$ average time. For $n$ objects, the average time complexity for computing $\rho$ which performs \textit{n} range queries in Algorithm~\ref{alg:qrho} is $\mathcal{O}(n \log_M{} n)$. Similarly, our modified query for $\delta$ in Algorithm~\ref{alg:qdelta} which explores nodes in a way
similar to nearest neighbor search but with additional pruning also takes $\mathcal{O}(n \log_M{} n)$ time for $n$ objects. Therefore, the overall average time complexity for DPC algorithm is $\mathcal{O}(n \log_M{} n)$.
\end{IEEEproof}

\section{Experimental Study} \label{exp}

We conduct a comprehensive set of experiments to evaluate the performance of all the indices (List, CH Index, Quadtree, R-tree). The evaluation is comprised of the following tasks:
(i) Query performance on different datasets with different sizes.
(ii) Preprocessing and storage costs of indices.
(iii) Query performance under influence of different parameters $d_c$, $bin width$ $w$, $neighbor\ threshold$ $\tau$.
(iv) Clustering quality.
Based on these tasks, experiments were conducted on a machine equipped with core i5 2.40 GHz processor, 16 GB of RAM and Windows 10 Operating system. All algorithms including the original DPC algorithm have been programmed in C++ and compiled using g++ and optimization level set to O3. We consider the original DPC algorithm as the baseline solution with which the efficiency of proposed algorithms and clustering quality of approximate solution are compared. 

\begin{table}
\centering
\renewcommand{\arraystretch}{0.9}
\caption{\\Summary of Datasets}\label{tb:sgt}
%\vspace{-5pt}
\resizebox{0.75\columnwidth}{!}{%
\begin{tabular}{|p{2.0cm}|p{1.5cm}| p{1.6cm}|}
\hline
\textbf{Dataset} & \textbf{Objects}  & \textbf{Types}\\
\hline 
S1\cite{franti2006iterative} & 5000  & Synthetic\\
Query\cite{savva2018explaining},\cite{anagnostopoulos2018scalable} & 50000  & Synthetic\\
Birch\cite{zhang1997birch} & 100000  & Synthetic\\
Range\cite{savva2018explaining},\cite{anagnostopoulos2018scalable} & 200000  & Synthetic\\
Brightkite\cite{cho2011friendship} & 399100 & Real\\
Gowalla\cite{cho2011friendship} & 1256680  & Real\\
\hline
\end{tabular}
}
\vspace{-5pt}
\end{table}

\textit{\textbf{Datasets.}} 
We used both synthetic and real datasets to evaluate the indices. The datasets, shown in Table~\ref{tb:sgt}, of different sizes
were used to test the algorithms.
S1 and Birch are benchmark datasets obtained from \cite{ClusteringDatasets} where S1 dataset contains around 5000 objects and 15 clusters while the Birch dataset contains around 100000 objects and 100 clusters. 
 The Query and Range datasets consist of around 50,000 and 200,000 objects respectively with spatial attributes. These have been obtained from UCI machine learning archive \cite{Dua:2019}. 
Brightkite and
Gowalla (available at SNAP~\cite{snapnets}) are real datasets of 399100 and 1.25 million unique user check-ins collected from their social network website. 

\textit{\textbf{Parameters.}} For each dataset, we inspected the query performance of algorithms for different $d_c$. 
For the list-based indices, different values of $\tau$ were chosen at large intervals to find the influence on running time, memory as well as clustering quality. The large intervals were needed to demonstrate the substantial effect on memory as well as running time. The influence of bin width \textit{w} was also analysed on the performance of CH Index. 

\textit{\textbf{Evaluation Metric.}} 
In order to analyse the clustering quality of our approximate solution, we use the well known measures Precision, Recall and F1 Score. 

\subsection{Running Time} 
\begin{figure}
\centering
\includegraphics[width=2.8in,height=1.38in]{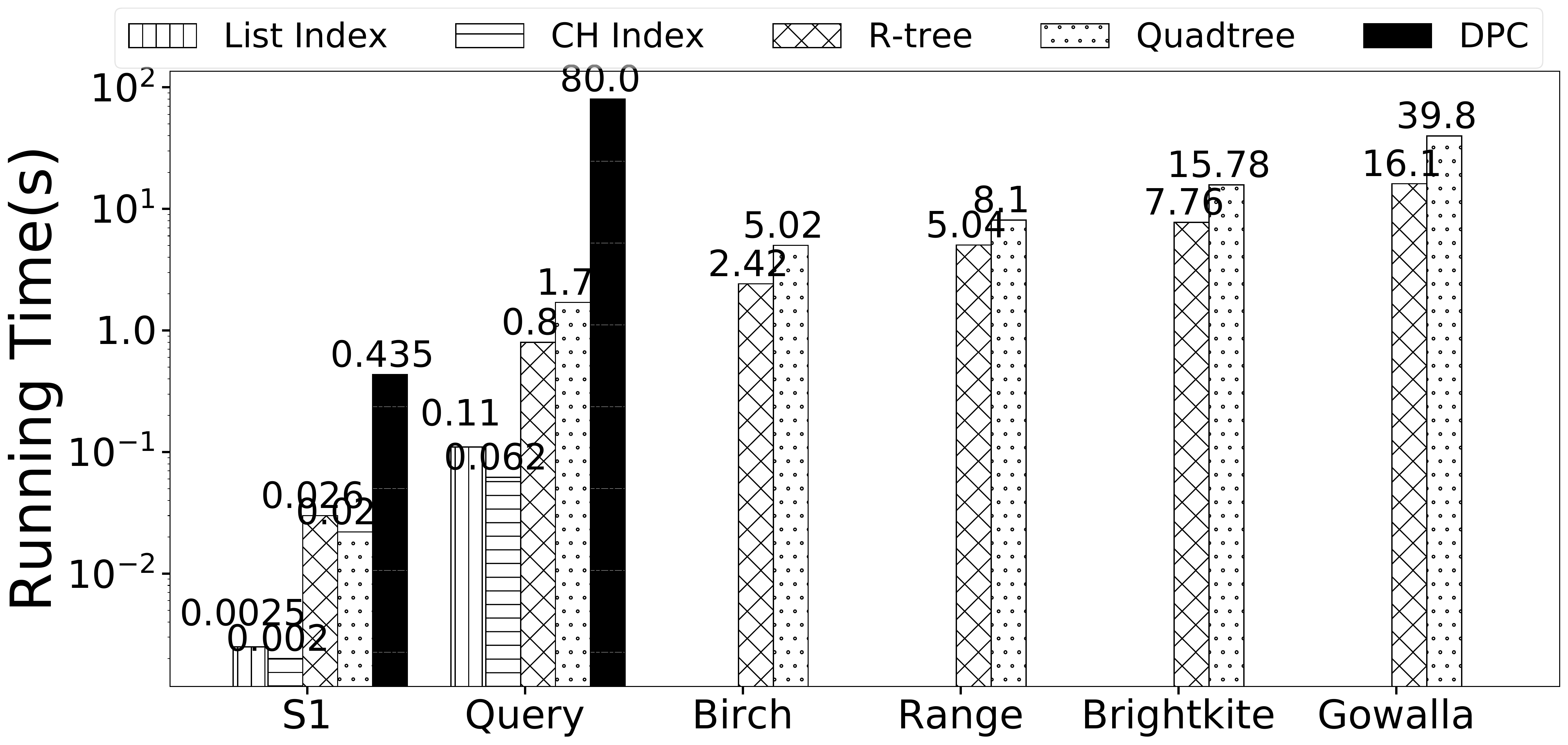}
\caption{Running Time comparison on various datasets} \label{fig:qtimedatasets}
\vspace{-5pt}
\end{figure}

We test the running time of all the algorithms on various datasets including the original DPC algorithm. Figure~\ref{fig:qtimedatasets} shows a bar graph where datasets are shown on the x-axis in the order of their non-decreasing sizes while the running time is shown on y-axis. For each dataset, comparisons were made for the same value of $d_c$. 

It is clear that for the small and medium datasets, S1 and Query, the list-based indices outperform the tree-based indices in providing the set of $\rho$ and $\delta$ values while CH Index performs the best. 
However, the original DPC took much more time. The running time of two datasets clearly show that with increase in size, DPC requires higher time. For the larger datasets, both DPC and list-based indices need large memory, and thus could not be tested. So, for the datasets Birch, Range, Brightkite and Gowalla, the graph does not contain any charts for the list-based indices and the original DPC algorithm. Therefore, the list-based indices with approximation were used to handle larger datasets, the results of which will be shown in Section~\ref{sec:appq}.

In general, with increase in size, the running time of all the algorithms increased. As to the tree-based indices, R-tree performs better than Quadtree on all synthetic and real datasets (except S1, where they exhibit similar performance) as shown in the figure. 
Faster query time of R-tree on large datasets is a result of its  balanced structure as compared to Quadtree. 

\begin{table}
\centering
\renewcommand{\arraystretch}{1.0}
\caption{\\Memory Usage by different Indices (in MB)}\label{tb:memusage}
%\vspace{-5pt}
\resizebox{0.90\columnwidth}{!}{%
\begin{tabular}{|p{1.4cm}|p{1.4cm}|p{1.3cm}|p{1.0cm}|p{1.3cm}|}
\hline
\textbf{Dataset} & \textbf{List Index} & \textbf{CH Index} & \textbf{R-tree} & \textbf{Quadtree}\\
\hline 
S1  & 98.7 & 99.1 & 5.2 & 9.6\\
Query  & 9501.8 & 9540.5 & 8.8 & 14.2\\
Birch  & 7004.1* & 7060.8* & 15.4 & 29.2\\
Range  & 8649.9* & 8720.4* & 28.6 & 55.9\\
Brightkite & 10367.0* & 10433.2* & 76.4 & 120.8 \\
Gowalla  & 7805.8* & 7875.2* & 140.2 & 244.8\\
\hline 
\end{tabular}
}
\vspace{-5pt}
\end{table}

\subsection{Memory Usage and Construction Time}\label{sec:preprocessing}
Table~\ref{tb:memusage} and Table~\ref{tb:const} show the memory requirement and construction time of different indices for various datasets. 
The `*' sign indicates the memory and construction time of the List Index and CH Index for the largest $\tau$ used. For Birch, Range, Brightkite and Gowalla datasets, $\tau$ is $250000$, $2500$, $1.0$ and $0.05$ respectively.
Selection of these $\tau$ has been done to fully utilize the memory. For CH Index, a small size bin width is selected for each of the above datasets whose values are 2000, 0.0006, 8000, 600, 0.02 and 0.015 respectively.

List-based indices clearly require significantly large storage space compared to tree-based indices. Among them, CH Index requires slightly extra cost to store the cumulative histograms. However, experiments show that using a small RN-List, the memory consumption can be controlled. For example, although, Birch, Range, Brightkite and Gowalla have different sizes, the index constructed have nearly the same size. Further, R-tree needs slightly lower memory than Quadtree (owing to its balanced structure).

The index construction time is considerably less for tree-based indices as compared to list-based indices as shown in Table~\ref{tb:const}. For CH Index, we only report the extra time taken to build histograms on top of List Index. The table shows that histograms do not add much burden on the construction cost as compared to List Index. The construction time of Quadtree is less than R-tree for the small and medium datasets because R-tree has to consider the balanced structure. However, for the larger Brightkite and Gowalla datasets, it took similar or more time implying that the balancing cost of R-tree is offset for large datasets.
The tree-based indices have extremely low memory consumption as well as fast preprocessing time as compared to the list-based indices which makes them superior in terms of index construction.

\begin{table}
\centering
\renewcommand{\arraystretch}{1.0}
\caption{\\Construction Time of different Indices (in Sec)}\label{tb:const}
%\vspace{-5pt}
\resizebox{0.90\columnwidth}{!}{%
\begin{tabular}{|p{1.4cm}|p{1.4cm}|p{1.3cm}|p{1.0cm}|p{1.3cm}|}
\hline
\textbf{Dataset}  & \textbf{List Index} & \textbf{CH Index} & \textbf{R-tree} & \textbf{Quadtree}\\
\hline 
S1  & 15.230 & 6.821 & 0.015 & 0.004\\
Query  & 1580.596 & 96.140 & 0.290 & 0.040 \\
Birch  & 1282.690* & 52.370* & 0.250 & 0.046\\
Range & 2078.069* & 68.460* & 0.340 & 0.124\\
Brightkite & 4001.148* & 254.521* & 0.442 & 0.432 \\
Gowalla  & 29070.08* & 69.150* & 0.925 & 1.946\\
\hline
\end{tabular}
}
\vspace{-5pt}
\end{table} 

\subsection{Effect of Parameter Variation} \label{sec:parv}
This section examines the effect of different parameters on the behaviour of different query algorithms.

\subsubsection{Effect of $d_c$}
\begin{figure}
\centering
\subfloat[S1]{\includegraphics[width=1.70in,height=1.3in]{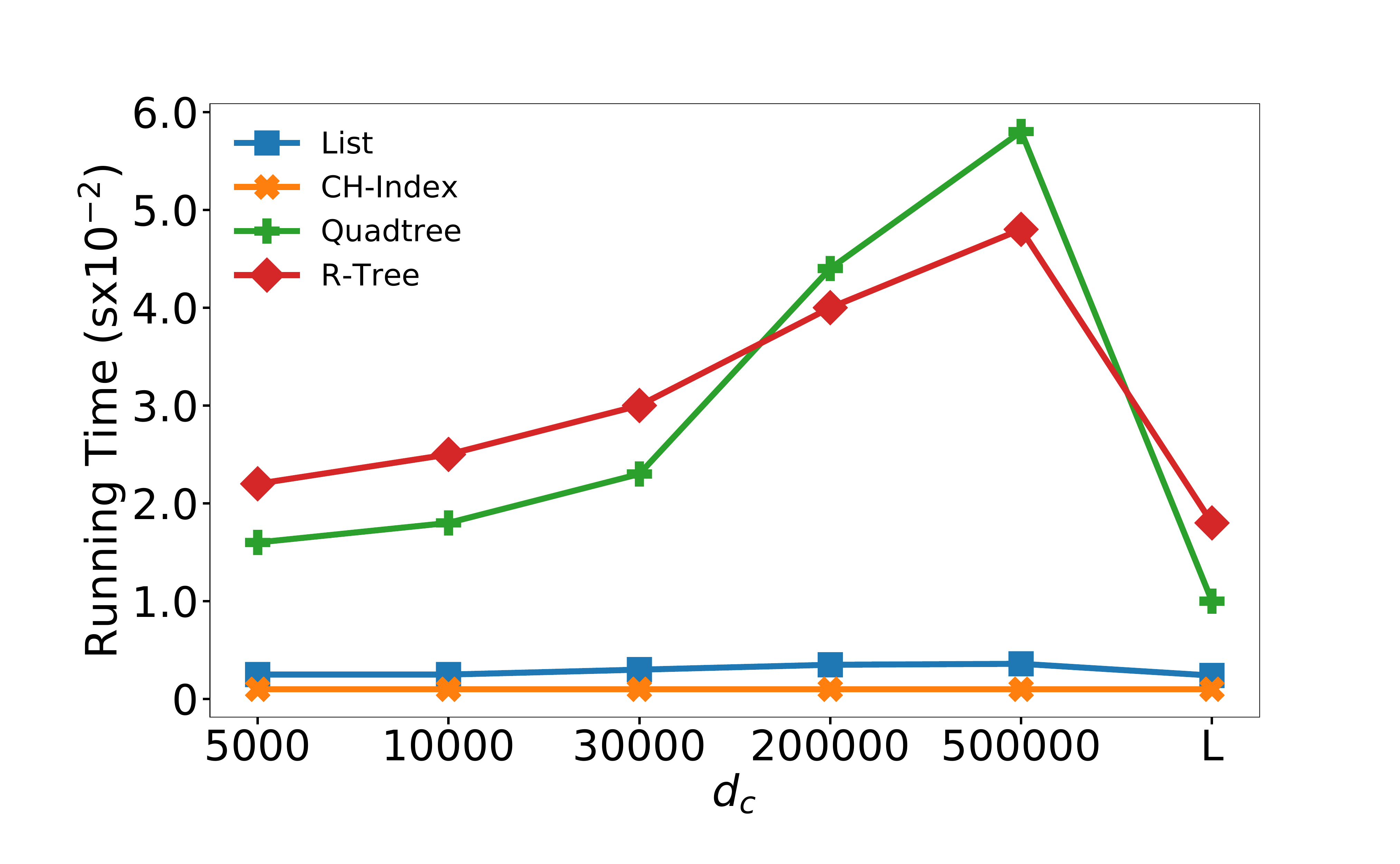}}
\hspace{1pt}
\subfloat[Query]{\includegraphics[width=1.70in,height=1.3in]{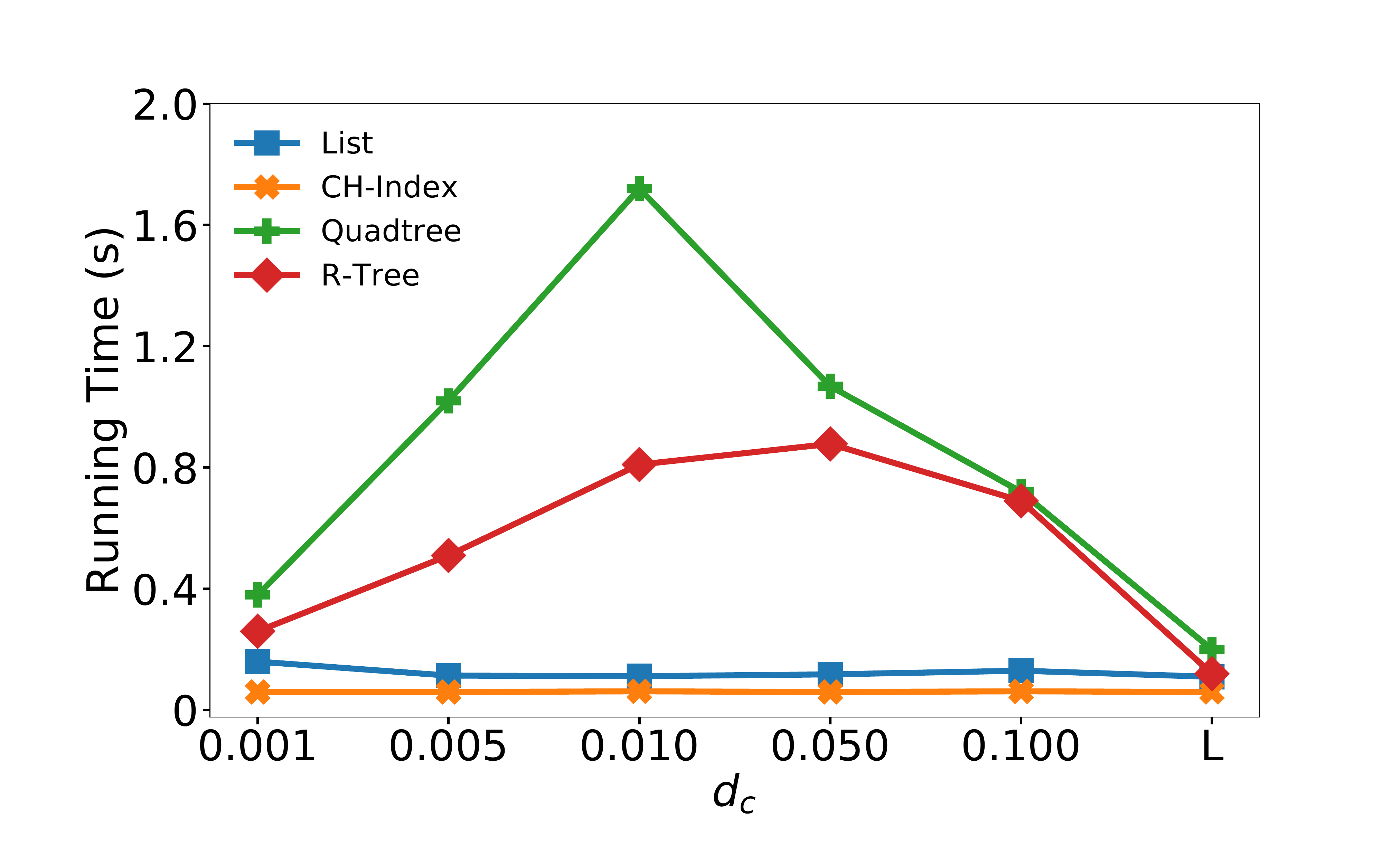}}
\vspace{-1pt}
\subfloat[Birch]{\includegraphics[width=1.70in,height=1.3in]{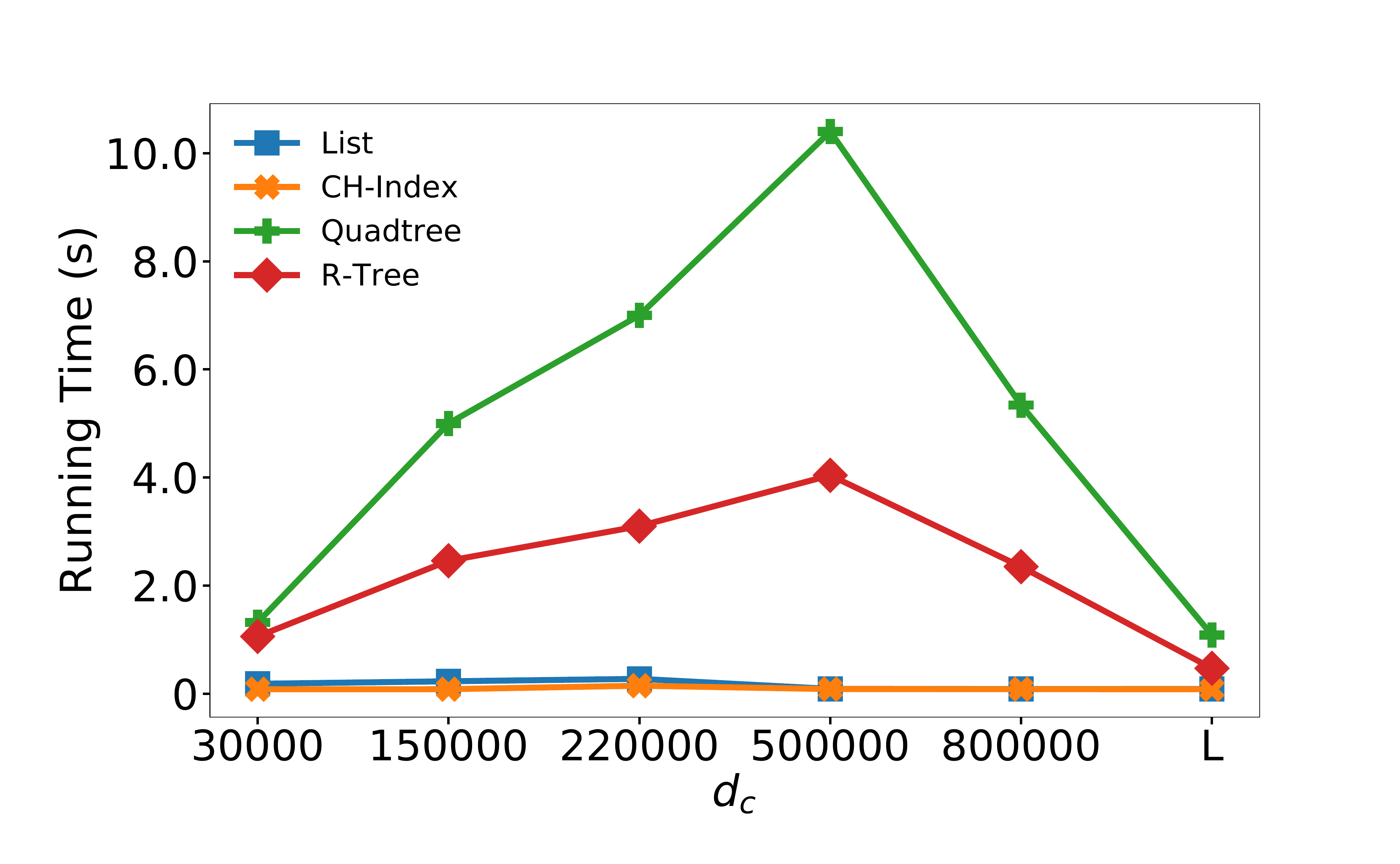}}
\label{fig:chamdc}
\hspace{1pt}
\subfloat[Range]{\includegraphics[width=1.70in,height=1.3in]{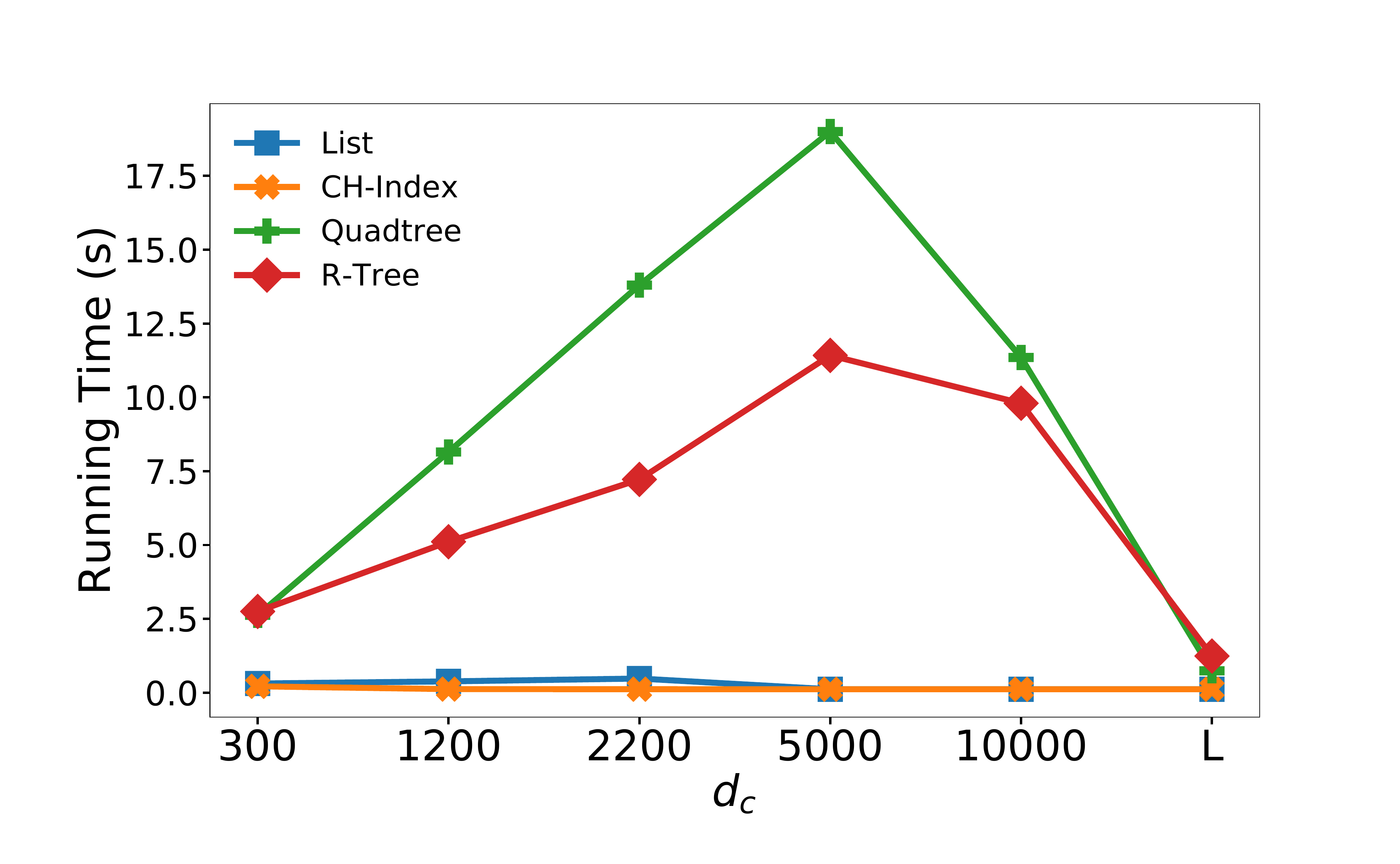}}
\label{fig:birchdc}
\subfloat[Brightkite]{\includegraphics[width=1.70in,height=1.3in]{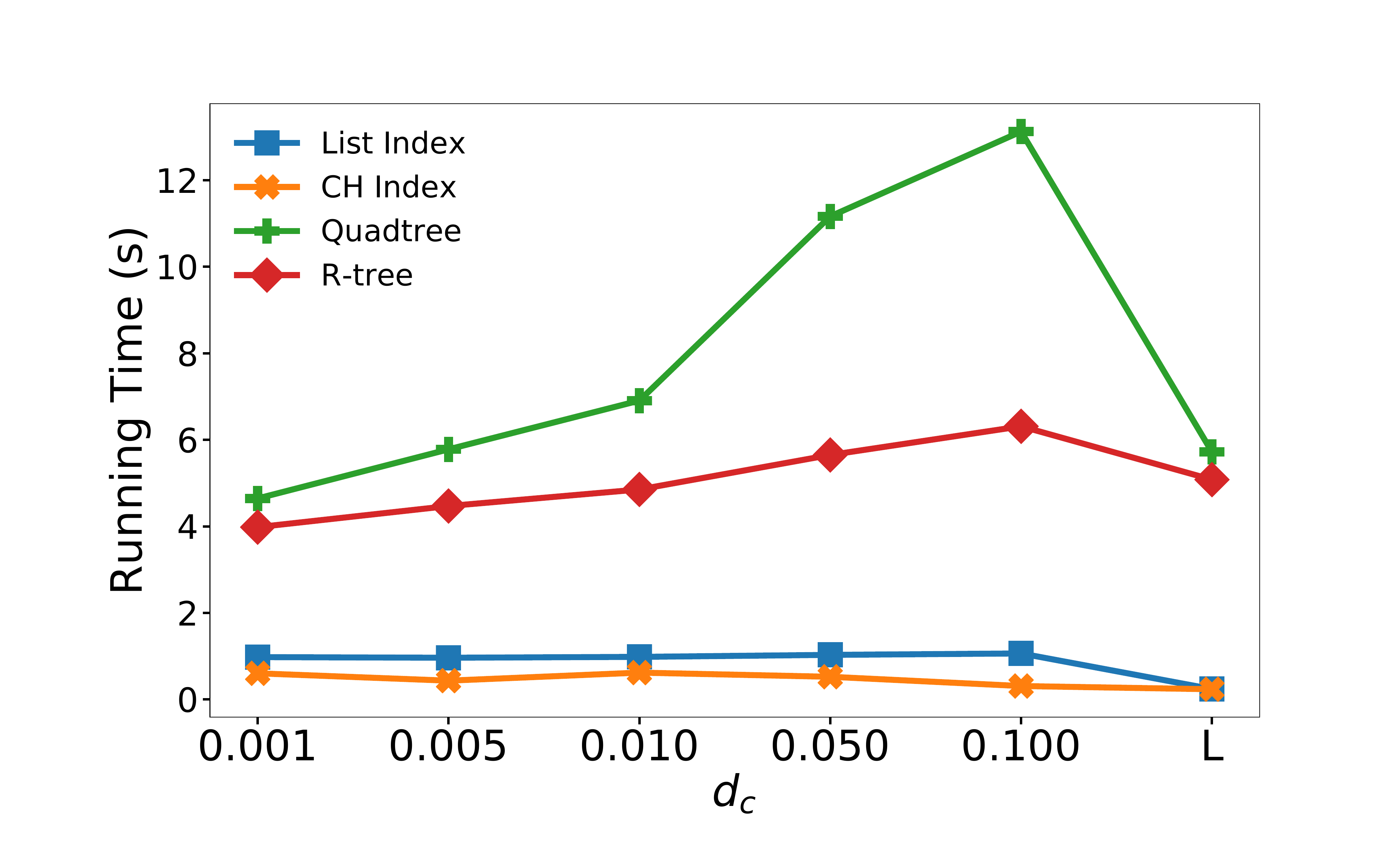}}
\hspace{1pt}
\subfloat[Gowalla]{\includegraphics[width=1.70in,height=1.3in]{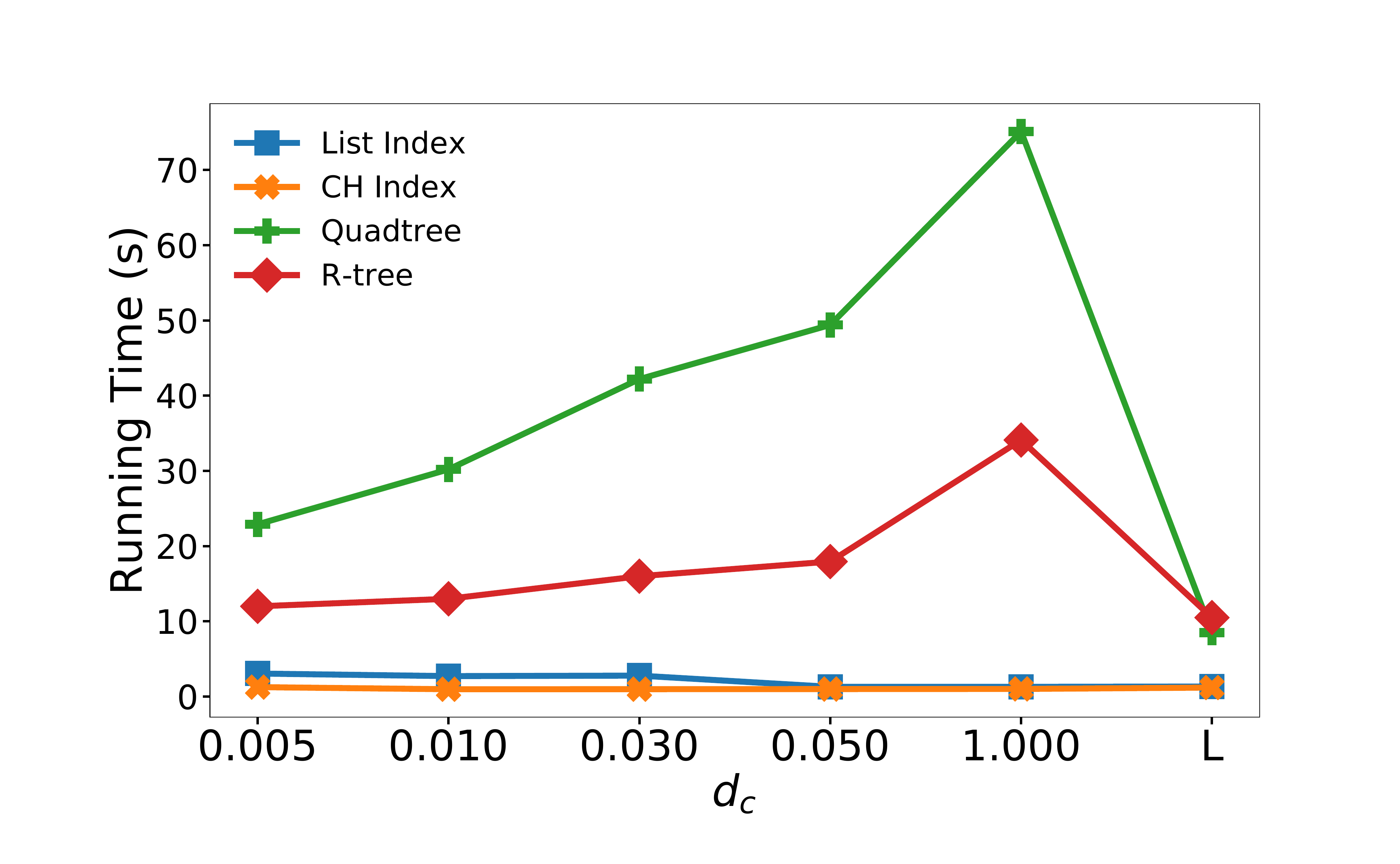}}
\label{fig:godc}
\caption{Running Time vs $d_c$}\label{fig:runt}
\vspace{-15pt}
\end{figure}

Figure~\ref{fig:runt} shows the influence of $d_c$ on the running time of algorithms on different datasets. The values of $d_c$ depend on the underlying space of each dataset. For the Birch, Range, Brightkite and Gowalla, we used the largest $\tau$ as in the previous set-up. The results of list-based indices show no significant difference in running time for all datasets with varying $d_c$. 
However, if $d_c>\tau$, the running time drops at the expense of loss of accuracy as no search operations are performed for the computation of $\rho$.

The running time of tree-based indices generally increases with increasing $d_c$ as more nodes are explored and more distance computations are performed during the computation of $\rho$. However, for large $d_c$, the general trend does not hold and the running time significantly drops from the expected behaviour. This is because of the pruning we developed, which avoids exploring most of the tree nodes as they are within $d_c$ from the queried object. For the largest $d_c$, $\rho$ for an object is equal to the total number of objects which is obtained in constant time as the root's \textit{nc} is directly assigned to $\rho$. As all the indices do not perform any search operation for largest $d_c$ and directly obtain $\rho$, their running times are very close as shown in the figure.

\subsubsection{Effect of Bin Width}
\begin{figure*}

        \centering
        \subfloat[Birch]{\includegraphics[width=1.7in,height=1.3in]{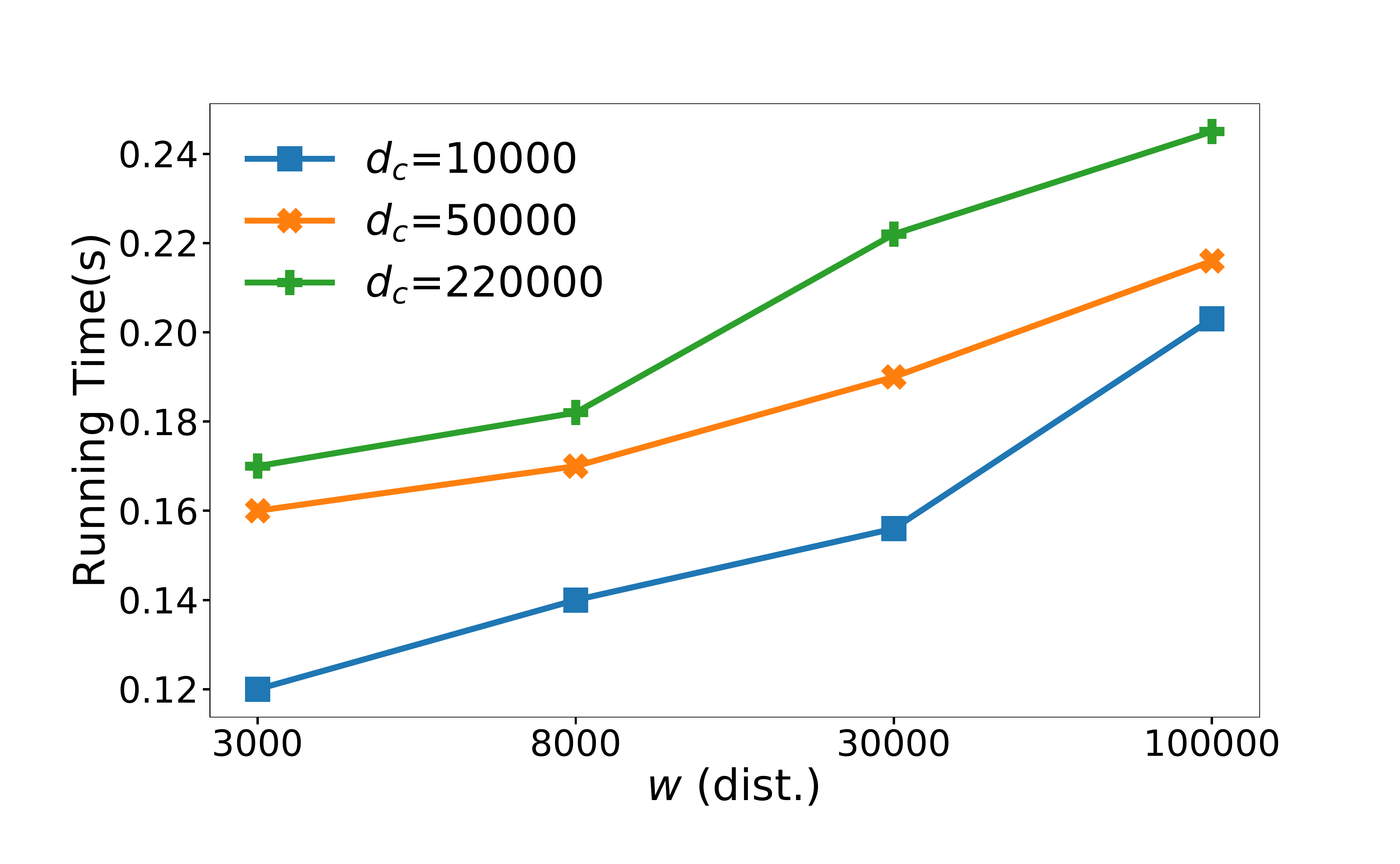}}
        \label{fig:birchbin}
        \hspace{1pt}
        \centering
        \subfloat[Range]{\includegraphics[width=1.7in,height=1.3in]{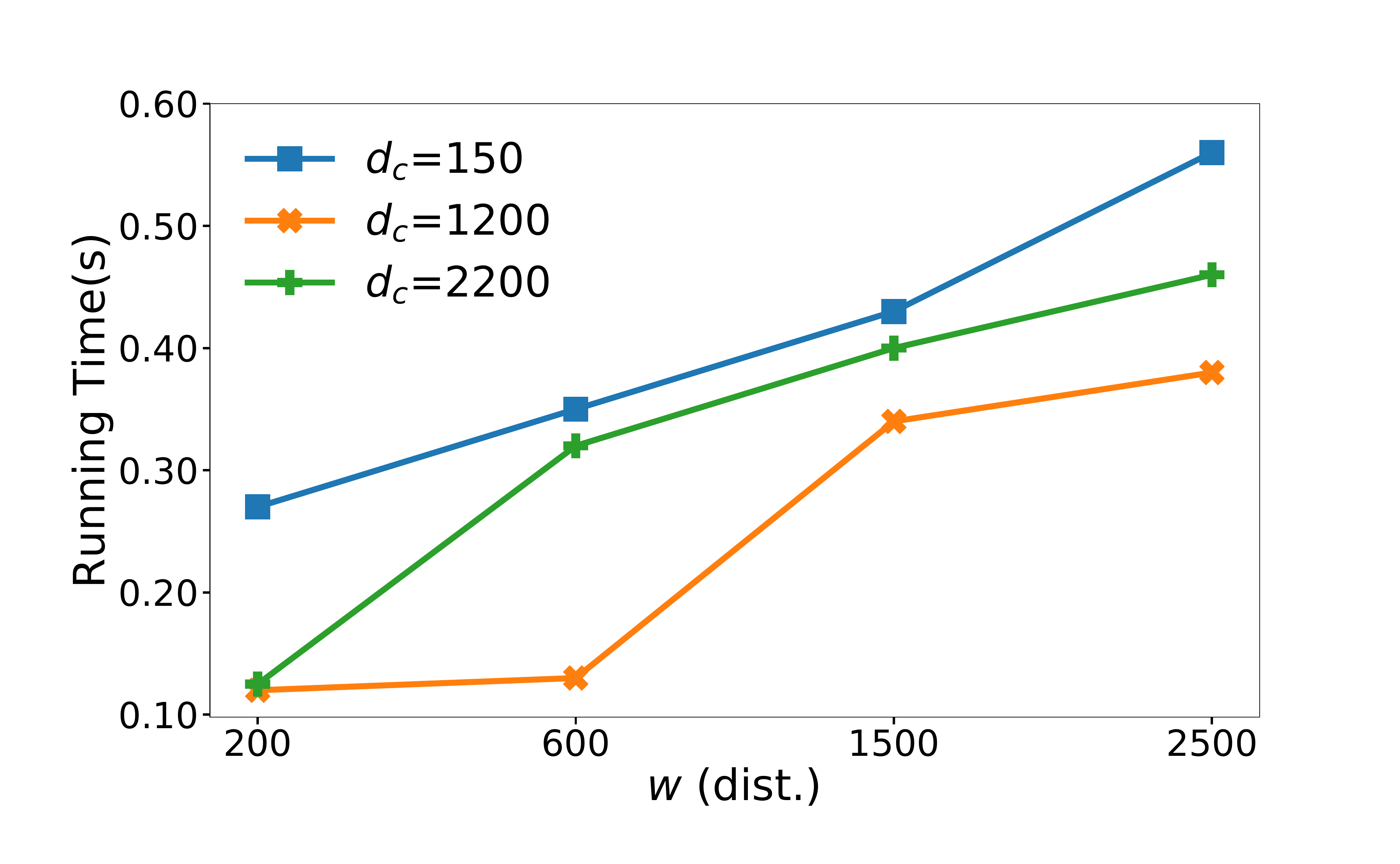}
        \label{fig:rangebin}}
        \hspace{1pt}
        \centering
        \subfloat[Brightkite]{\includegraphics[width=1.7in,height=1.3in]{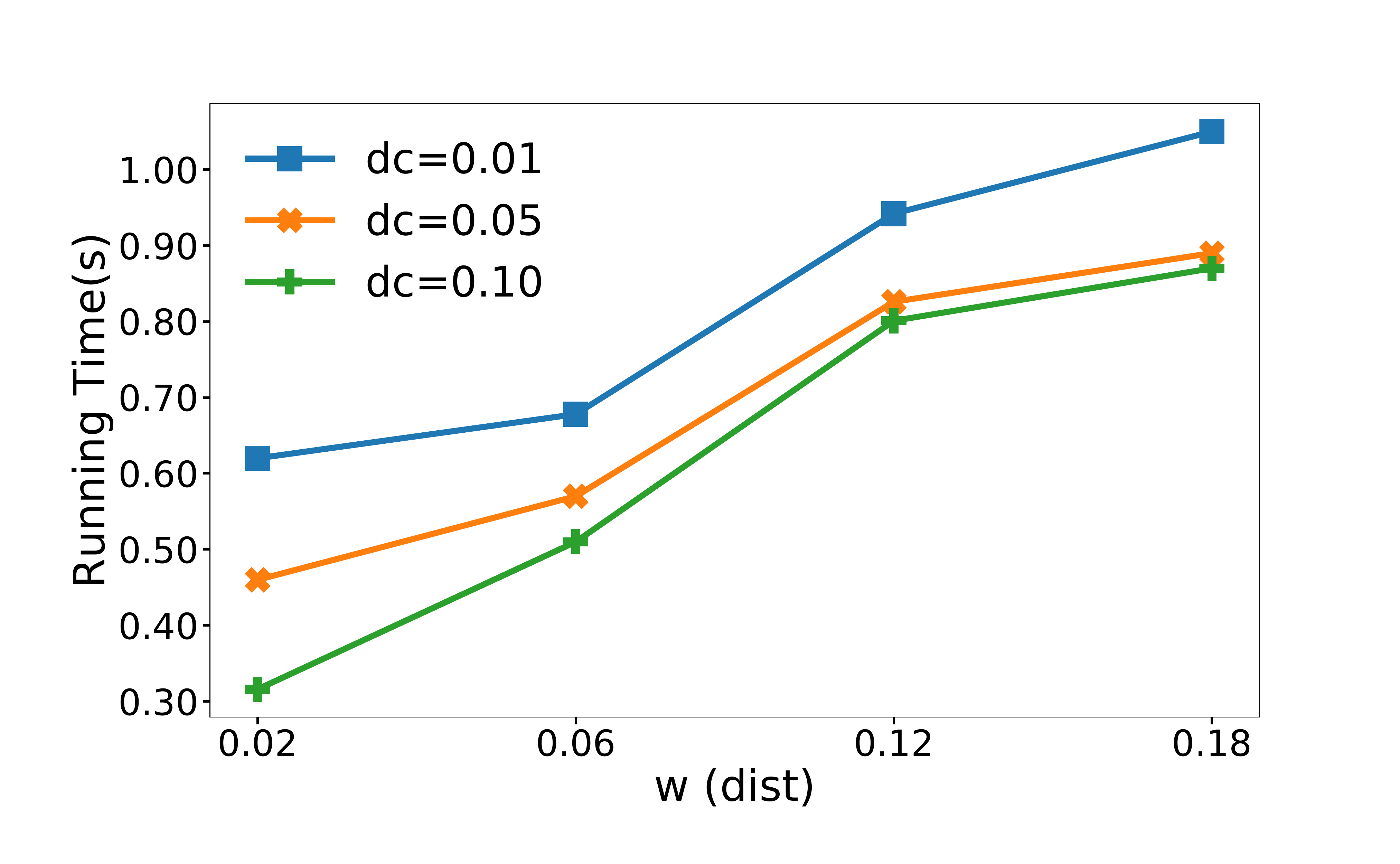}}
        \label{fig:bktbin}
        \hspace{1pt}
        \centering
        \subfloat[Gowalla]{\includegraphics[width=1.7in,height=1.3in]{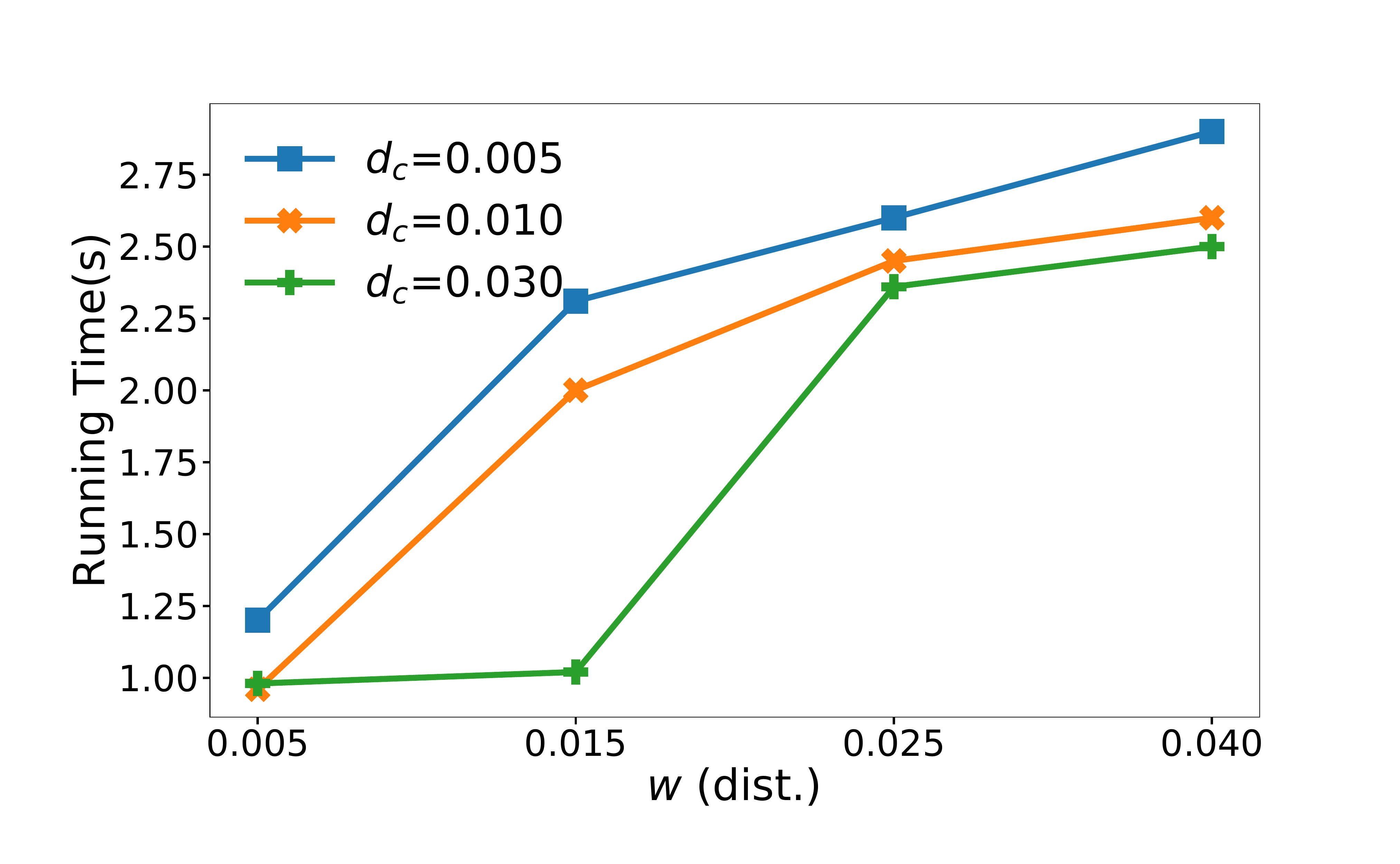}\label{fig:gowallabin}}
        %\vspace{-2pt}
        \caption{Influence of $w$}
        \label{fig:varyBin}
        %\vspace{-1pt}
        \hspace{1pt}
        \centering
        \subfloat[Birch]{\includegraphics[width=1.7in,height=1.3in]{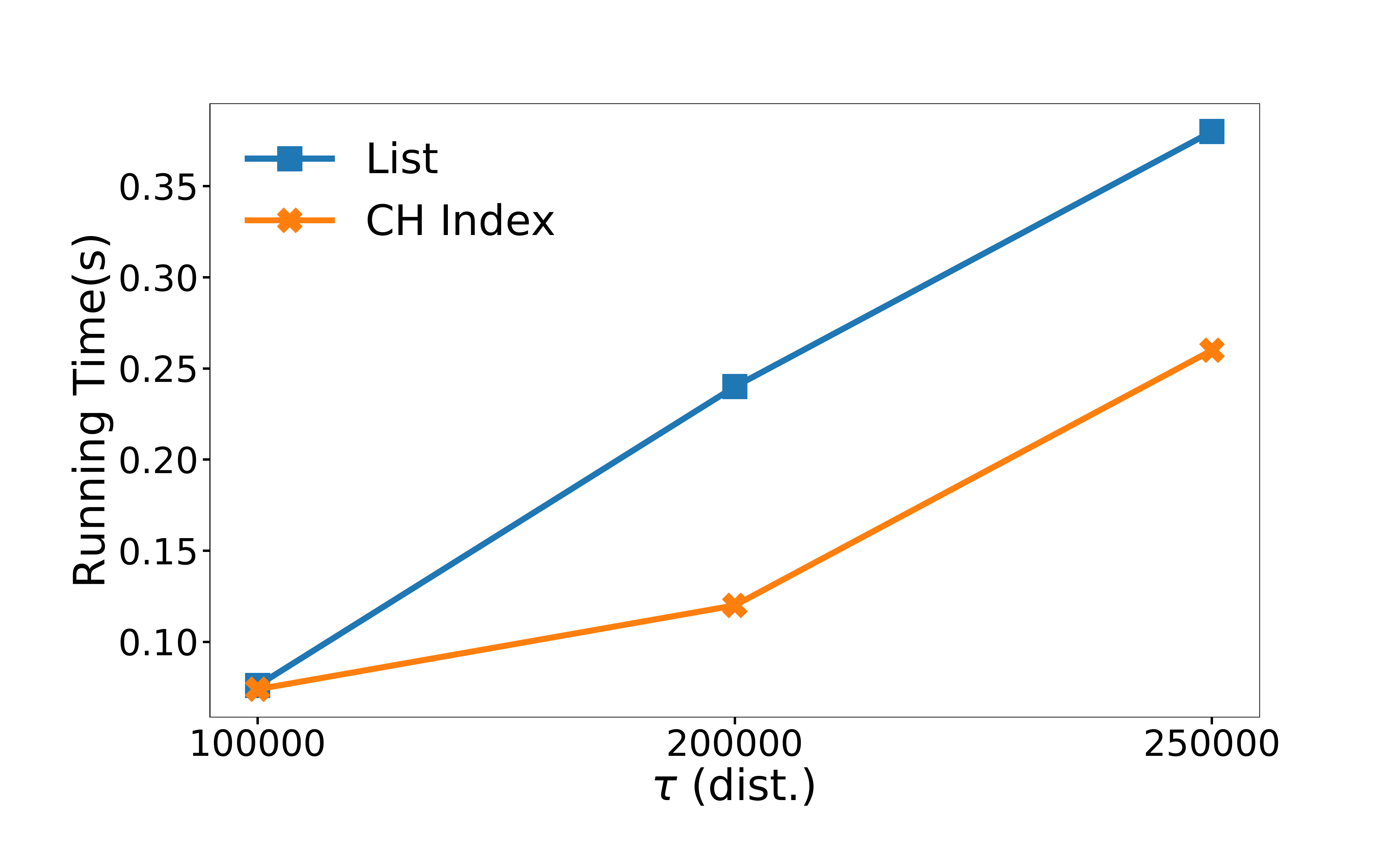}}
        \label{fig:birchtau}
        \hspace{1pt}
        \centering
        \subfloat[Range]{\includegraphics[width=1.7in,height=1.3in]{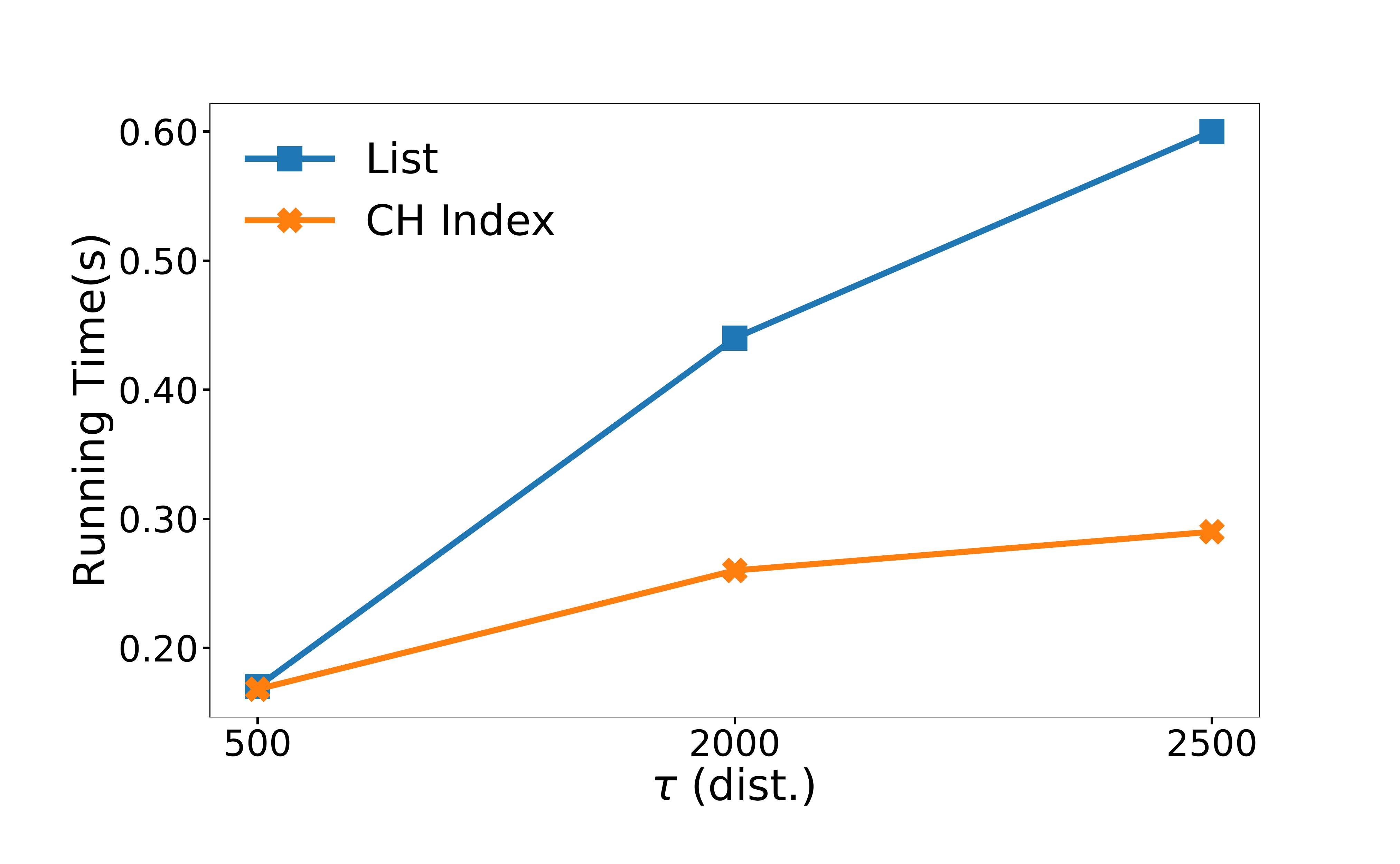}}
        \label{fig:rangetau}
        \hspace{1pt}
        \centering
        \subfloat[Brightkite]{\includegraphics[width=1.7in,height=1.3in]{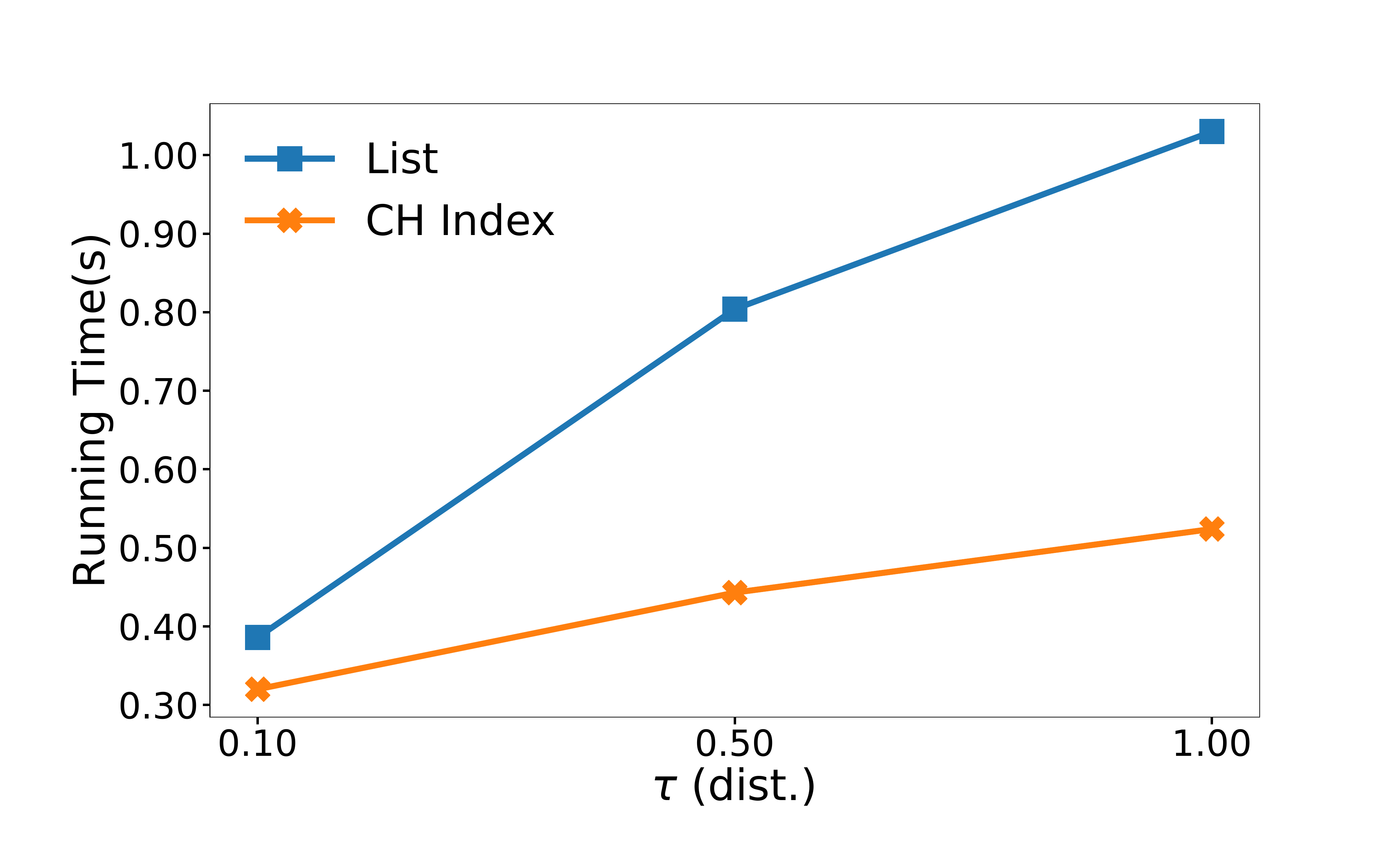}}
        \label{fig:bkttau}
        \hspace{1pt}
        \centering
        \subfloat[Gowalla]{\includegraphics[width=1.7in,height=1.3in]{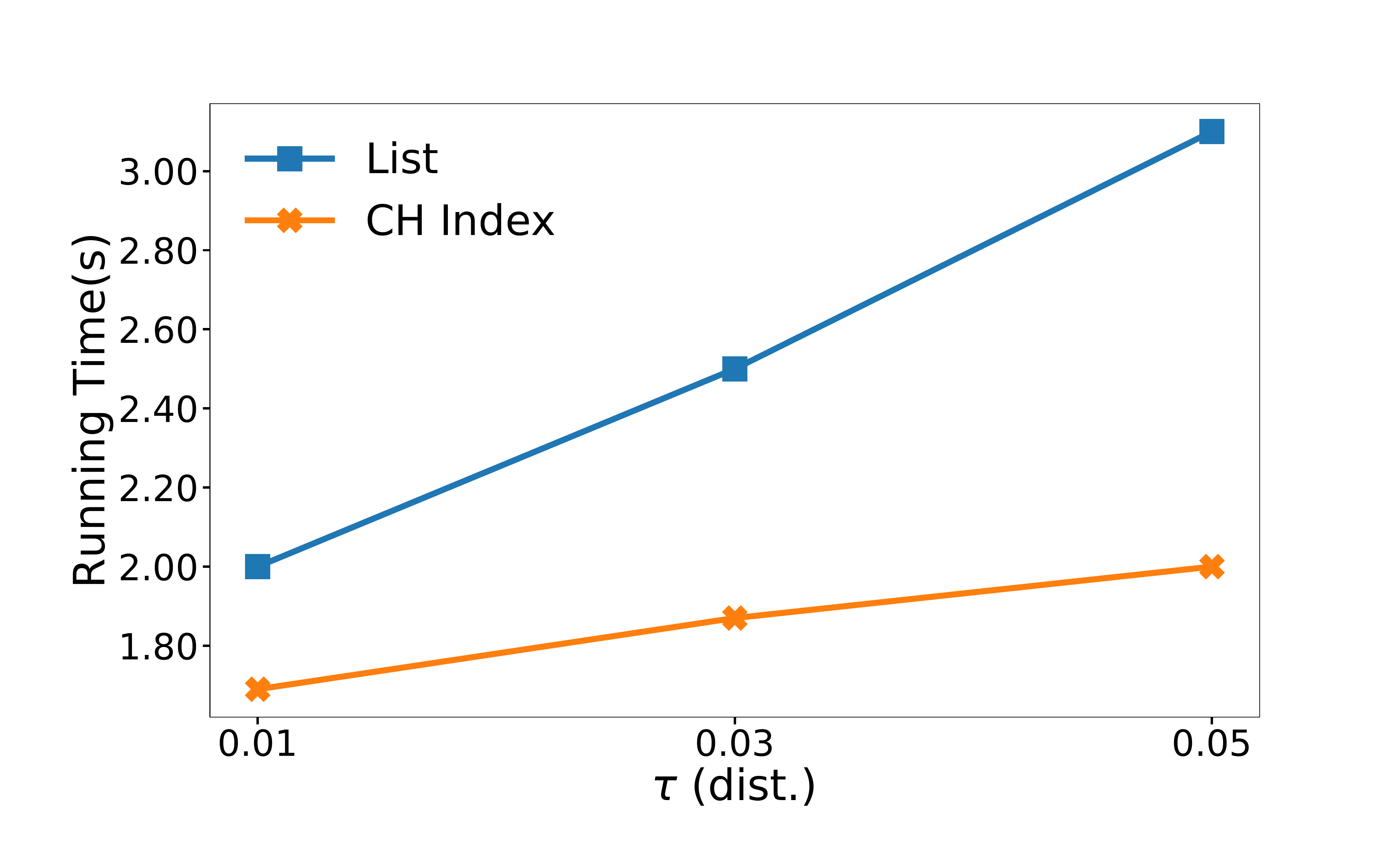}}
        \label{fig:gowallatau}
        %\vspace{-2pt}
        \caption{Influence of $\tau$}
        \label{fig:varyTau}
        \vspace{-15pt}
\end{figure*}

We also analysed the influence of \textit{bin width(w)} on the running time of CH Index for the Birch, Range, Brightkite and Gowalla datasets as shown in Figure~\ref{fig:varyBin}. Four different values of \textit{w} are selected and tested for three different values of $d_c$. Large intervals were chosen to show the significant differences in results. Note that, all datasets have different sets of \textit{w}, which depends on their sizes of region.

The results show that running time, in general, follows linear growth with increase in \textit{w}. Larger the value of \textit{w}, greater the running time. When \textit{w} is large, a large section of list needs to be searched, thereby taking more time. A slight deviation from general trend may occur in certain cases (for e.g., $d_c=1200$ in Figure ~\ref{fig:rangebin}, $d_c=0.030$ in Figure ~\ref{fig:gowallabin}), if $d_c$ is equal to a multiple of $\textit{w}$, the bin density gets directly assigned to $\rho$, for all objects, thereby improving the running time.

\begin{figure}[t]
\centering
\vspace{-10pt}
\subfloat[\textit{w} vs Memory]{{\includegraphics[width=1.70in,height=0.9in]{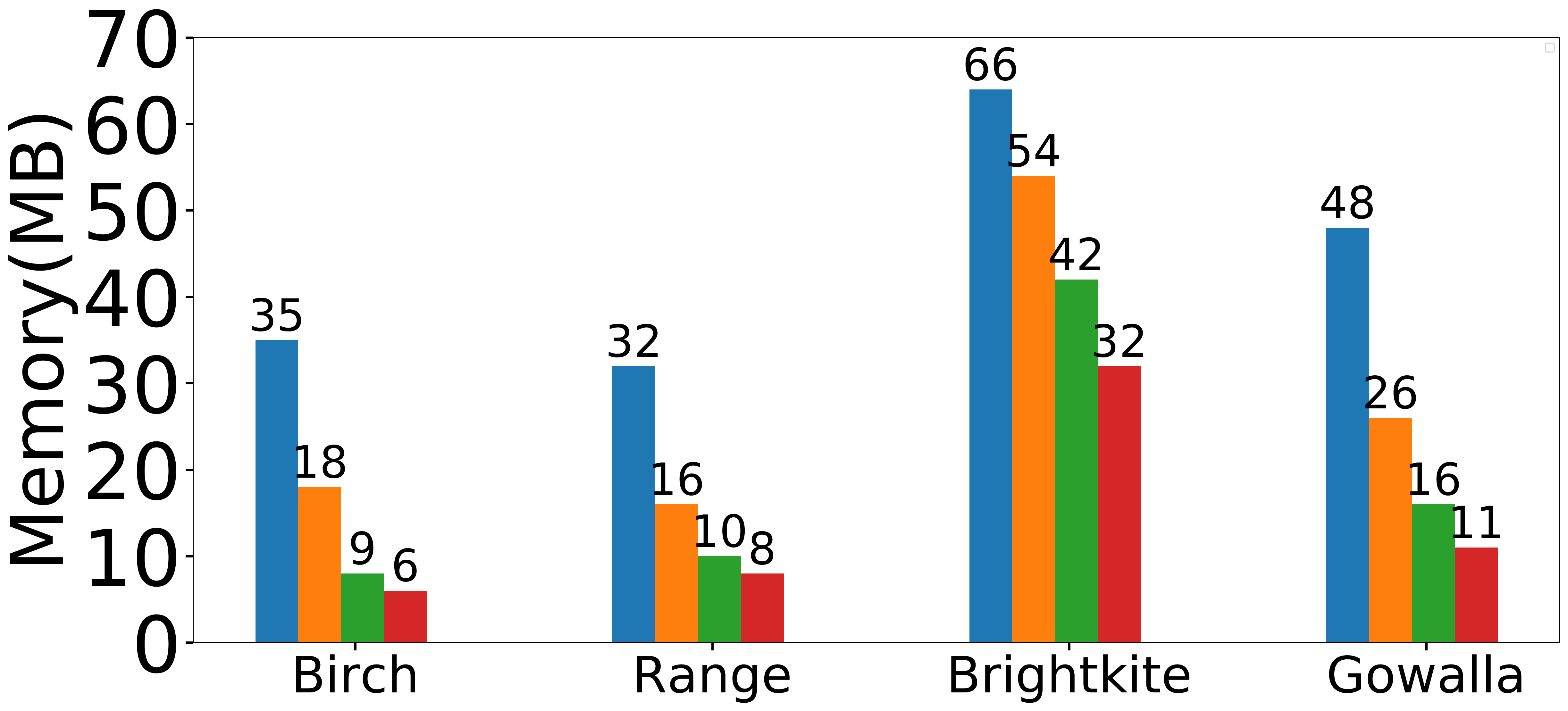}}
\label{fig:binM}}
\hfill
\centering
\subfloat[\textit{$\tau$} vs Memory]{{\includegraphics[width=1.70in,height=0.9in]{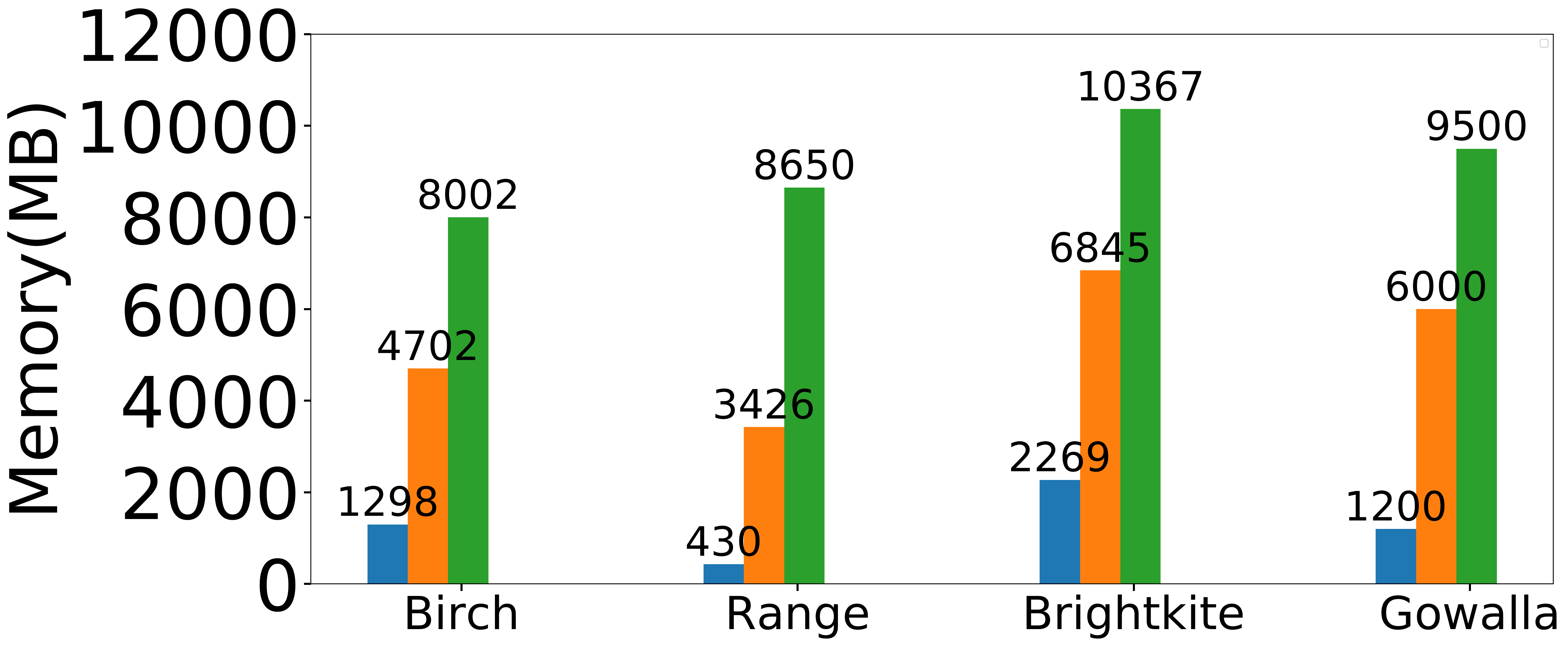}}
\label{fig:tauM}}
\vspace{-1pt}
\caption{Influence of \textit{w} and \textit{$\tau$} on Memory}\label{fig:mem}
\vspace{-10pt}
\centering
\subfloat[Birch dataset]{{\includegraphics[width=1.7in,height=1.3in]{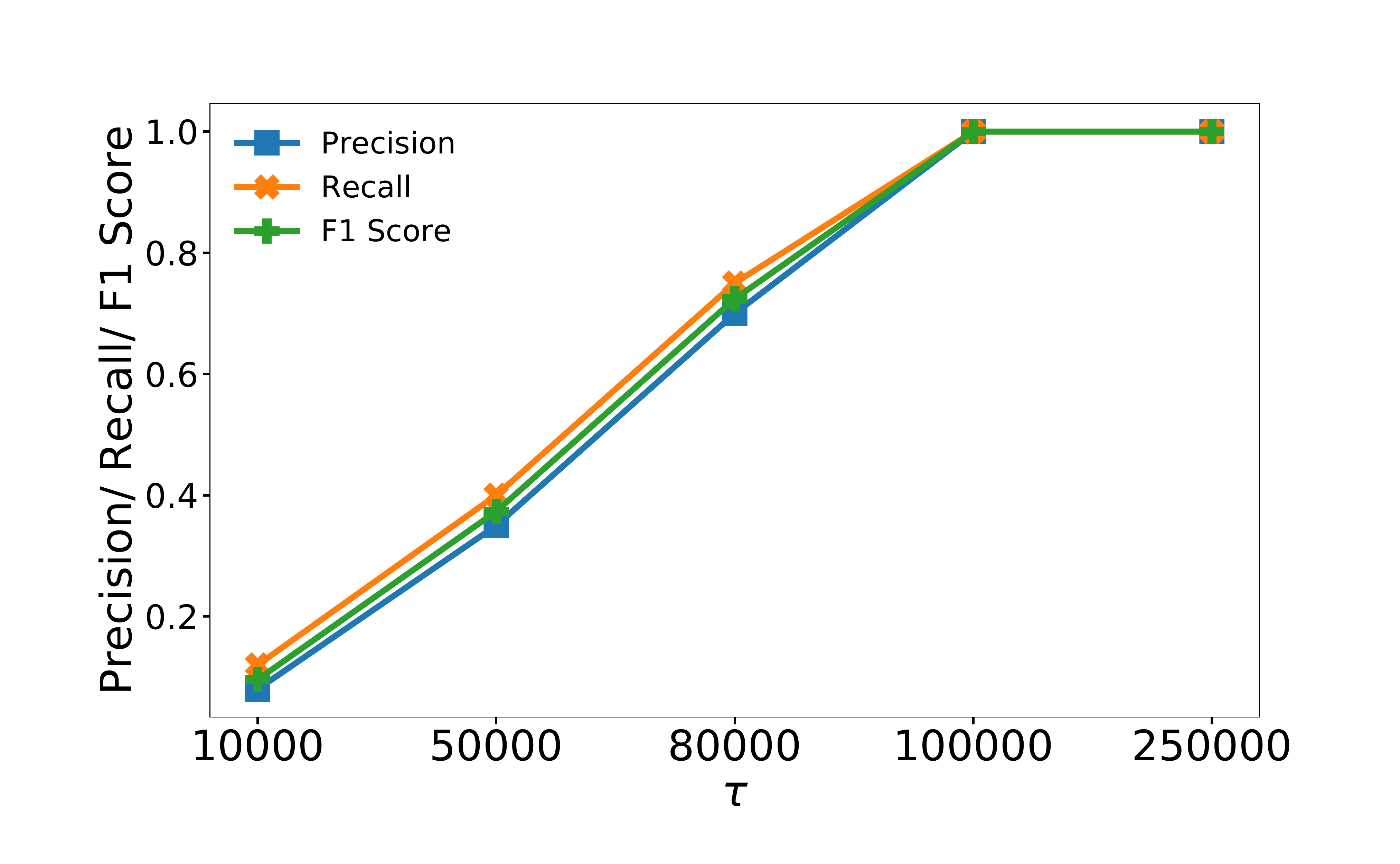}}
\label{fig:cq1}}
\hfill
\centering
\subfloat[Range dataset]{{\includegraphics[width=1.7in,height=1.3in]{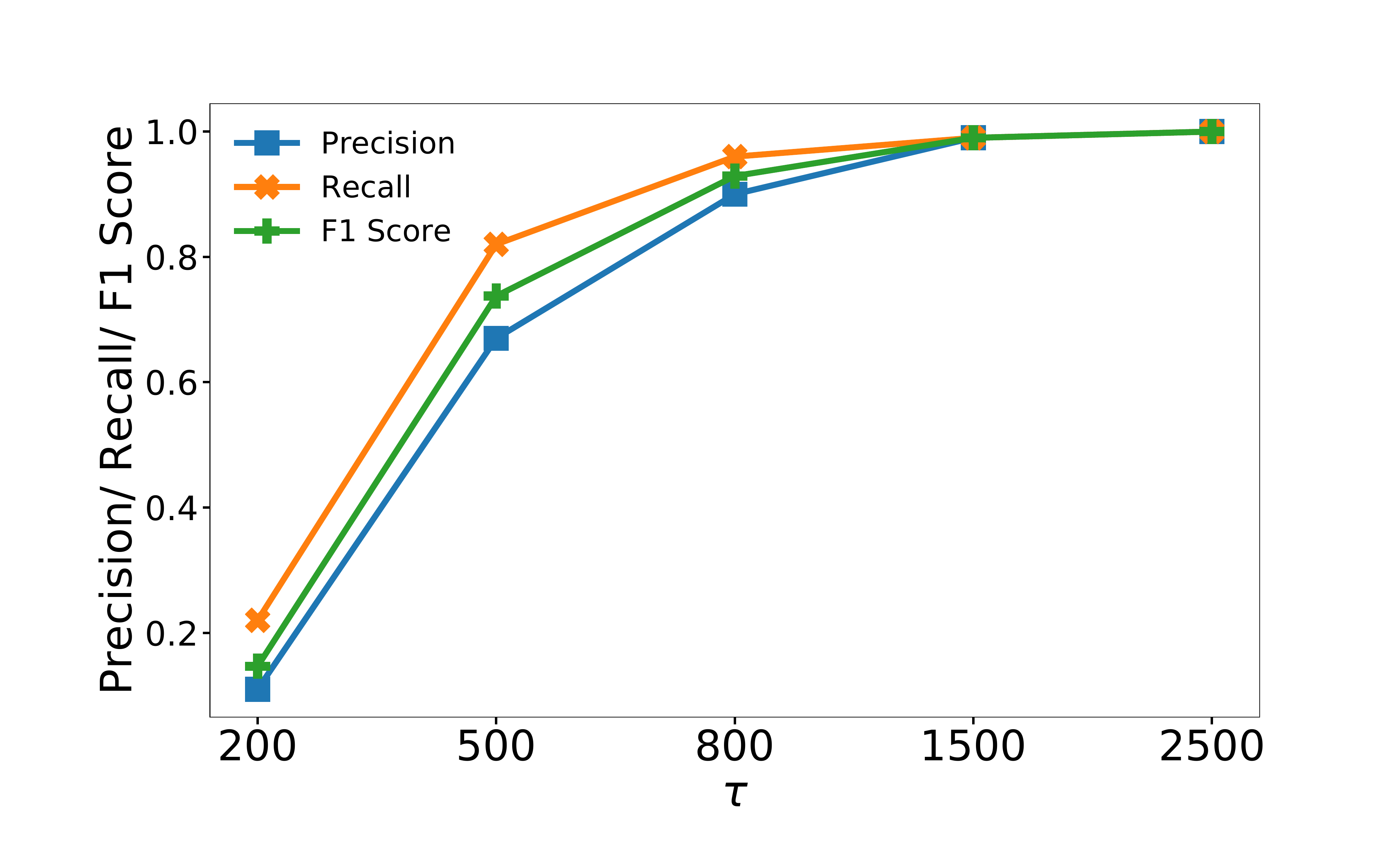}}
\label{fig:cq2}}
\vspace{-12pt}
\centering
\subfloat[Brightkite dataset]{{\includegraphics[width=1.7in,height=1.3in]{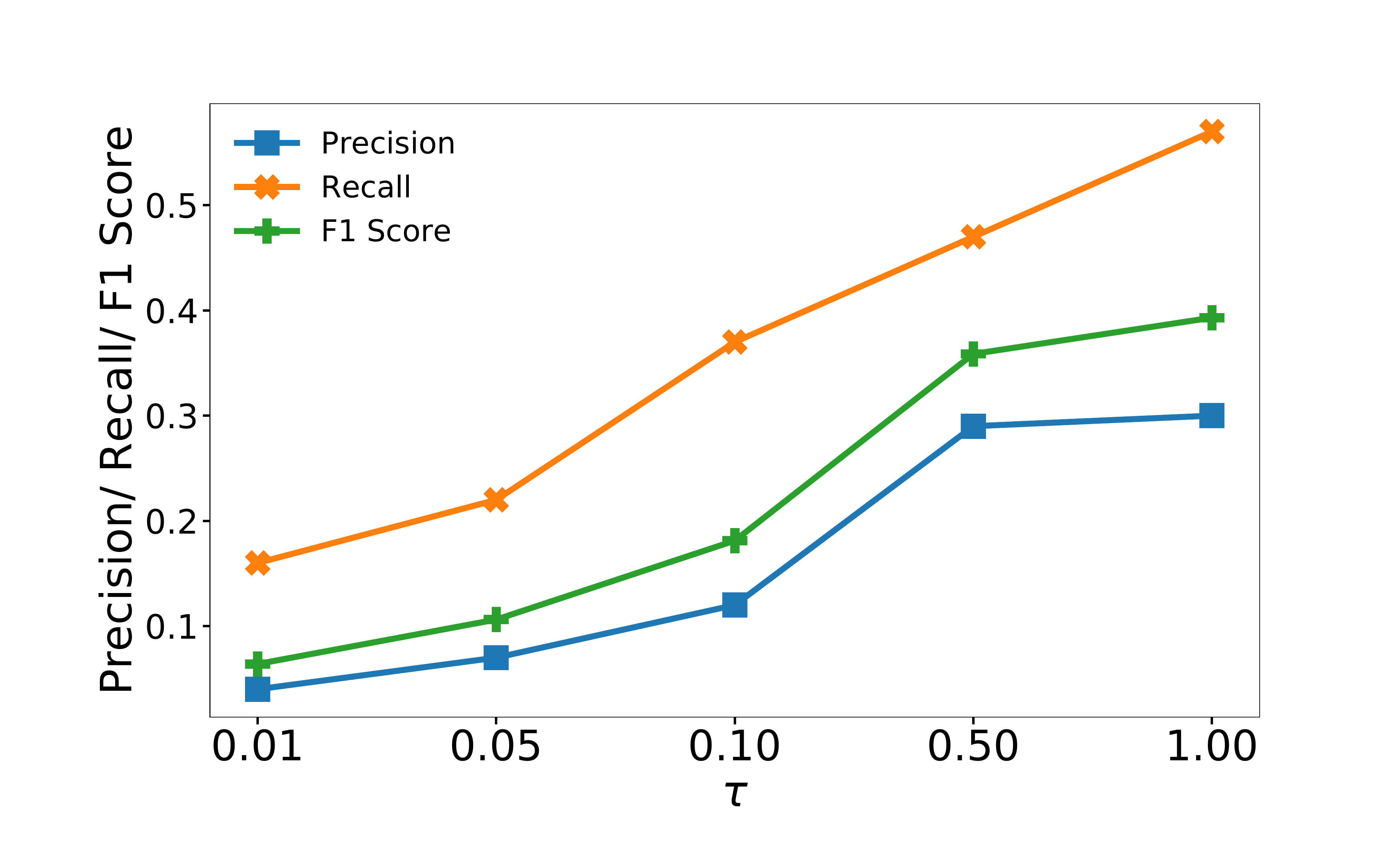}}
\label{fig:cq3}}
\hfill
\centering
\subfloat[Gowalla dataset]{{\includegraphics[width=1.7in,height=1.3in]{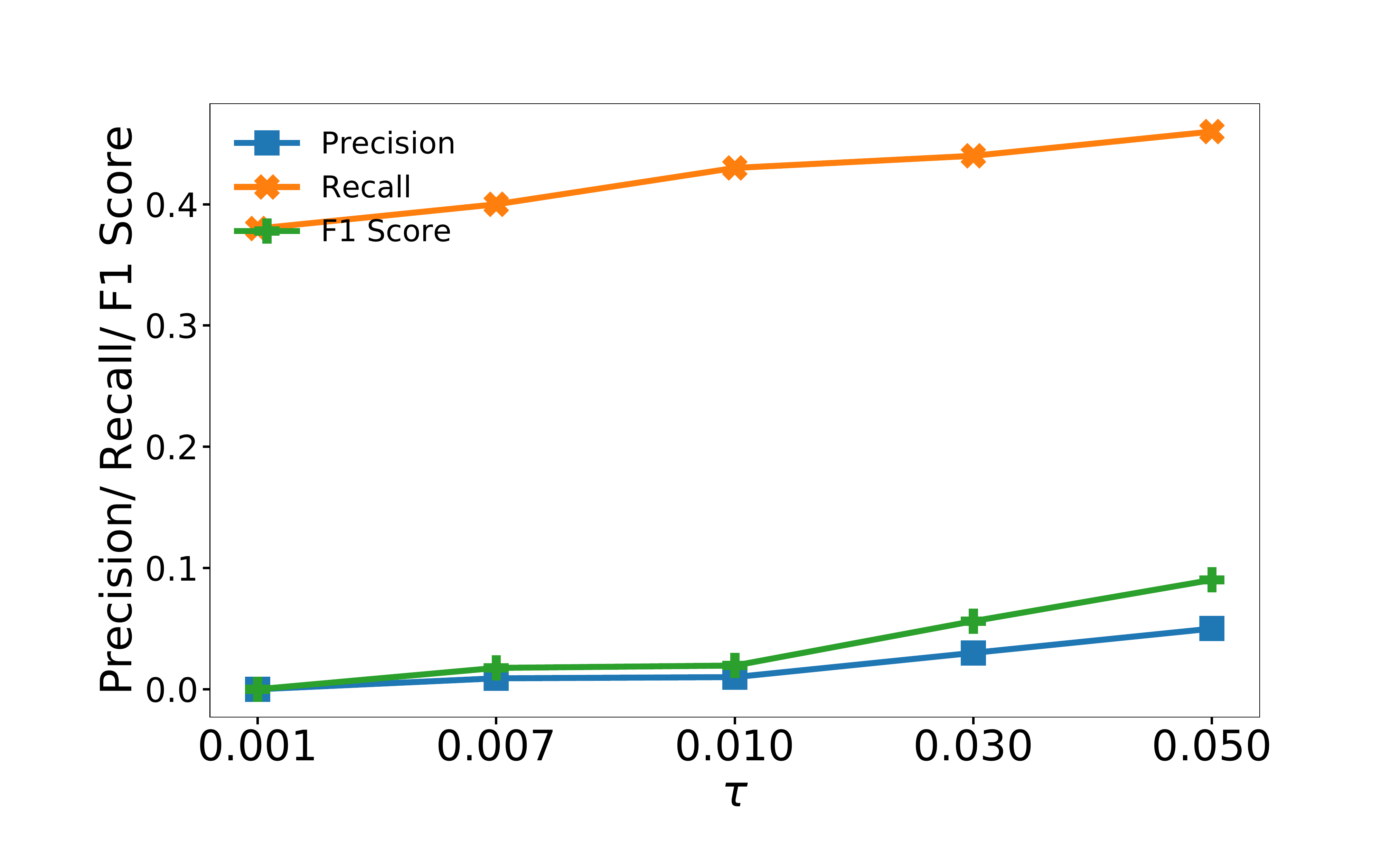}}
\label{fig:cq4}}
\caption{Clustering Quality of List Index with varying $\tau$}\label{fig:cq}
\vspace{-15pt}
\end{figure}

In Figure~\ref{fig:binM}, the effect of varying \textit{w} on the memory required by a cumulative histogram of CH Index is shown for each dataset. With \textit{w} increased, the memory consumption is reduced because, for large \textit{w}, only few bins are stored for each object. Similarly, for smaller \textit{w}, a large number of bins are required, which requires more memory. Thus, a CH Index with smaller \textit{w} provides better running time at the cost of space and vice-versa.

\subsection{Approximate Index Evaluation}\label{sec:appq}
This section analyses the effect of approximation on the running time, memory and clustering quality of list-based indices for Birch, Range, Brightkite and Gowalla datasets for which complete Index could not be stored. The parameter $d_c$ is fixed at 100000, 1500, 0.5, 0.001 respectively.

Figure~\ref{fig:varyTau} shows the variation of running time with $\tau$. We selected three different values of $\tau$ greater than the selected $d_c$ for each dataset.  
The results clearly show that, for both List Index and CH Index, the running time is directly proportional to $\tau$ or the size of RN-List. The shorter the RN-List, the lower the running time and vice-versa, since binary search performs the query on smaller number of objects in the list. When $\tau$ is very small, their running time are comparable because of nearly same search space. Moreover, the running time variation with $\tau$ is smaller for CH Index, because for varying $\tau$, the differences in running time come from the computation of $\delta$ only, since they take almost the same time in computing $\rho$ (as \textit{w} is fixed).

Figure~\ref{fig:tauM} shows the influence of $\tau$ on memory of List Index for different datasets. Clearly, the larger the $\tau$, the higher will be the space cost and vice-versa. The selection of smaller $\tau$ provides faster running time and low memory requirements at the cost of accuracy.

Finally, we examine the quality of clustering results obtained for different $\tau$ w.r.t. the results of the original DPC algorithm. Additional $\tau$ values have been selected to give a better idea about the results. As mentioned earlier, we use Precision, Recall and F1 Score metrics for this purpose. These measures try to evaluate cluster membership of object pairs in the reference clustering $G$ and the obtained clustering $C$. For this purpose, clustering result of DPC is taken to be reference clustering $G$. Precision tries to find the proportion of pairs correctly identified in $C$ to the total number of pairs in $C$, while Recall reflects the proportion of correct pairs identified in $C$ to the total number of pairs in $G$. These are defined as follows:

\begin{equation}
Precision = \frac{TP}{TP + FP}
\end{equation}

\begin{equation}
Recall = \frac{TP}{TP + FN}
\end{equation}

Here, $TP$ denotes $True\ Positive$ which states the number of cases when a pair of objects found in $C$ is also found in $G$. $FP$ denotes $False\ Positive$ which states the number of cases when a pair of objects found in $C$ is not found in $G$. $FN$ denotes $False\ Negative$ stating the number of cases when a pair of objects in $G$ which does not appear in $C$. 

$F1$ Score tries to seek the balance between the two metrics Precision and Recall and is defined as the harmonic mean of both.

\begin{equation}
F1\ Score = \frac{2\times Precision \times Recall}{Precision + Recall}
\end{equation}

If $\tau$ is reduced, many objects may be assigned a wrong $\delta$. Thus, many true positives will be lost and many false positives and false negatives occur which will result in decrease of both precision and recall and consequently affect the F1 Score. Therefore, the higher the values of these metrics obtained, the better the clustering quality. 

Figure~\ref{fig:cq} shows the three metric values of the clustering obtained using list-based indices for different values of $\tau$.  
Note that, RN-List of list-based index using the selected largest $\tau$ is still smaller than the complete N-List. For each dataset, we fix $d_c$ and determine the metric values while reducing $\tau$. 
For the datasets in Figure~\ref{fig:cq1} and Figure~\ref{fig:cq2}, 
the results clearly indicate that almost correct clustering results were obtained when $d_c \leq \tau$. All the three metric values over 0.98 which signifies list-based indices found approximately the same clusters as the original DPC algorithm. If $\tau$ is further reduced slightly below $d_c$, the metric values indicate slight differences in results.
As $\tau$ is further reduced, the metrics fall dramatically to very low values indicating poor clustering results. 

Interestingly, for the selected $d_c$ and largest $\tau$, for which correct results were obtained, we found only 1\% of the index for Range dataset and only 3\% of the index for Birch dataset were probed by the queries, which signifies the strength of the approximate solution.
However, for the larger dataset, Brightkite and Gowalla, only a very small RN-List could be loaded into memory, which resulted in incorrect $\delta$ values for most objects. As such, the metric values dropped significantly with Precision and F1 Score dropped below 0.4 for Brightkite and below 0.05 for Gowalla as shown in Figure~\ref{fig:cq3} and Figure~\ref{fig:cq4}. This shows that the approximate solution can support only the small and medium size datasets but does not lend well to large datasets.

\subsection{Discussion}
In this section, the pros and cons of different indices have been compared and summarized. This aids the users to select an appropriate index.  
 The proposed List-based indices for DPC comprises two index structures: List Index and CH Index. 
List Index is a simple, yet efficient index structure. 
Fast response to queries is achieved at the high cost of storing neighbors for each object.  
Another drawback of List Index is the expensive computation of $\rho$ for large datasets.
CH Index remedies the second drawback of List Index by capturing the statistics of each list as cumulative histograms. These histograms show faster response time for $\rho$ queries. This improvement comes at the cost of some extra preprocessing time and memory cost.

 With lower memory requirement and better preprocessing cost, tree-based indices help to localize the queries for efficient processing. Quadtree 
 follows a simple construction approach, due to which, the preprocessing cost is generally less. The running time of Quadtree is higher than the list-based indices and R-tree because of the large number of nodes explored. Moreover, the structure of Quadtree may be unbalanced in some cases, which also contributes to its declining performance.
R-tree, on the other hand, has slightly higher preprocessing cost on small and medium datasets than Quadtree because it has to consider a balanced structure. However, this balancing results in better memory utilisation as well as improved running time. Next, we summarize the above comparison.

\medskip\noindent\textbf{Summary of Comparison.} List-based indices are fast methods to find DPC clusters and outperform tree-based indices w.r.t. the running time. 
With slight additional space, CH Index outperforms the List Index achieving 20-30\% improved running time. Therefore, for smaller datasets, we would recommend list-based indices, especially, CH Index. 
For medium datasets, if a user desires fast running time but allows slight approximation, list-based indices could still be a good choice. When dealing with large datasets, 
tree-based indices are recommended, since sufficient list-based indices cannot be loaded to memory even though the quality of results could be reasonably compromised.
Among the tree-based indices, R-tree is preferred which overall has better running time than Quadtree. Moreover, tree-based indices also dominate list-based indices in terms of index construction by large margins and can be preferred for such requirements.

\section{Related Works}
Several research works have focused on improving the efficiency, scalability and other aspects of the DPC method. 
Wu et al. \cite{wu2017fast} proposed a density and grid-based clustering method which avoids unnecessary pairwise distance computations. The method computes the density of grid nodes instead of object local densities using fuzzy-type approximation. 
A k-means based strategy was presented in \cite{bai2017fast} for enhancing the scalability of DPC. However, this method has high time complexity for large datasets. Accordingly, another approximate method was proposed to improve the speed based on exemplar clustering.
Xu et al. \cite{xu2018improved} proposed grid-based strategy to select dense grid cells and find the local density using the objects of those cells. To address the sparsity of grid cells, a circle division strategy was proposed. 
Zhang et. al. \cite{zhang2016efficient} proposed a distributed algorithm for DPC using MapReduce and employed locality sensitive hashing to obtain clustering results.

The works mentioned below have adapted DPC to deal with specific scenarios. In most cases, the density computation has been modified which is different from the original DPC. The work of Cheng et. al. \cite{cheng2019improved} deals with clustering specific manifold datasets.
Wang et al. \cite{wang2016automatic} proposed a new clustering algorithm which uses a density metric based on k-nearest neighbor (\textit{kNN}). The idea is that the dense objects will have \textit{kNN} very close as compared to the sparse objects. The local density was redefined using kNN. 
Moreover, a technique was also suggested for automatically detecting the cluster centers.
Chen et al. \cite{chen2018neighbourhood} proposed Neighbourhood Contrast (NC) as an alternative to density for detecting cluster centers which can admit all local density maxima. According to this, all local density maxima have similar NC values, irrespective of the density values. Zhu et al. \cite{zhu2018distance} proposed a multi-dimensional scaling method for identifying clusters with varied densities. Instead of scaling each attribute, it rescales the pairwise distance between objects.  

\section{Conclusion}
In this paper, we have studied index-based methods for density peak clustering. We have proposed two list-based indices namely List Index and CH Index for DPC. For the List Index, efficient algorithms have been proposed to compute the two DPC quantities for any $d_c$. The advanced CH Index further improves over List Index with faster running time. These indices have higher space requirement and therefore for memory-constrained systems, approximate solution has been suggested to reduce index space, if marginal variation of the clustering results is allowed. Moreover, two popular tree-based indices, Quadtree and R-tree, have also been studied for DPC to address the high space cost of list-based indices. Effective pruning techniques and efficient algorithms have been designed to enable these indices for computing DPC quantities. The experimental results demonstrate that i) CH Index outperforms other indices w.r.t. query time but suffers from high preprocessing and space cost, ii) approximate solution supports list-based indices to handle medium size datasets with near accurate clustering results, iii) R-tree is faster than Quadtree for DPC queries and suitable to handle medium to large datasets.

\section*{Acknowledgments}
This work is jointly supported by ARC Discovery Projects under Grant No. DP170104747, DP180100212, DP200103700, and National Natural Science Foundation of China under Grant No. 61872258. 

\ifCLASSOPTIONcaptionsoff
  \newpage
\fi

\bibliographystyle{IEEEtran}
\bibliography{IEEEabrv,bare_jrnl_compsoc}

\vspace{-24pt}
\begin{IEEEbiography}[{\includegraphics[width=1in,height=1.25in,clip,keepaspectratio]{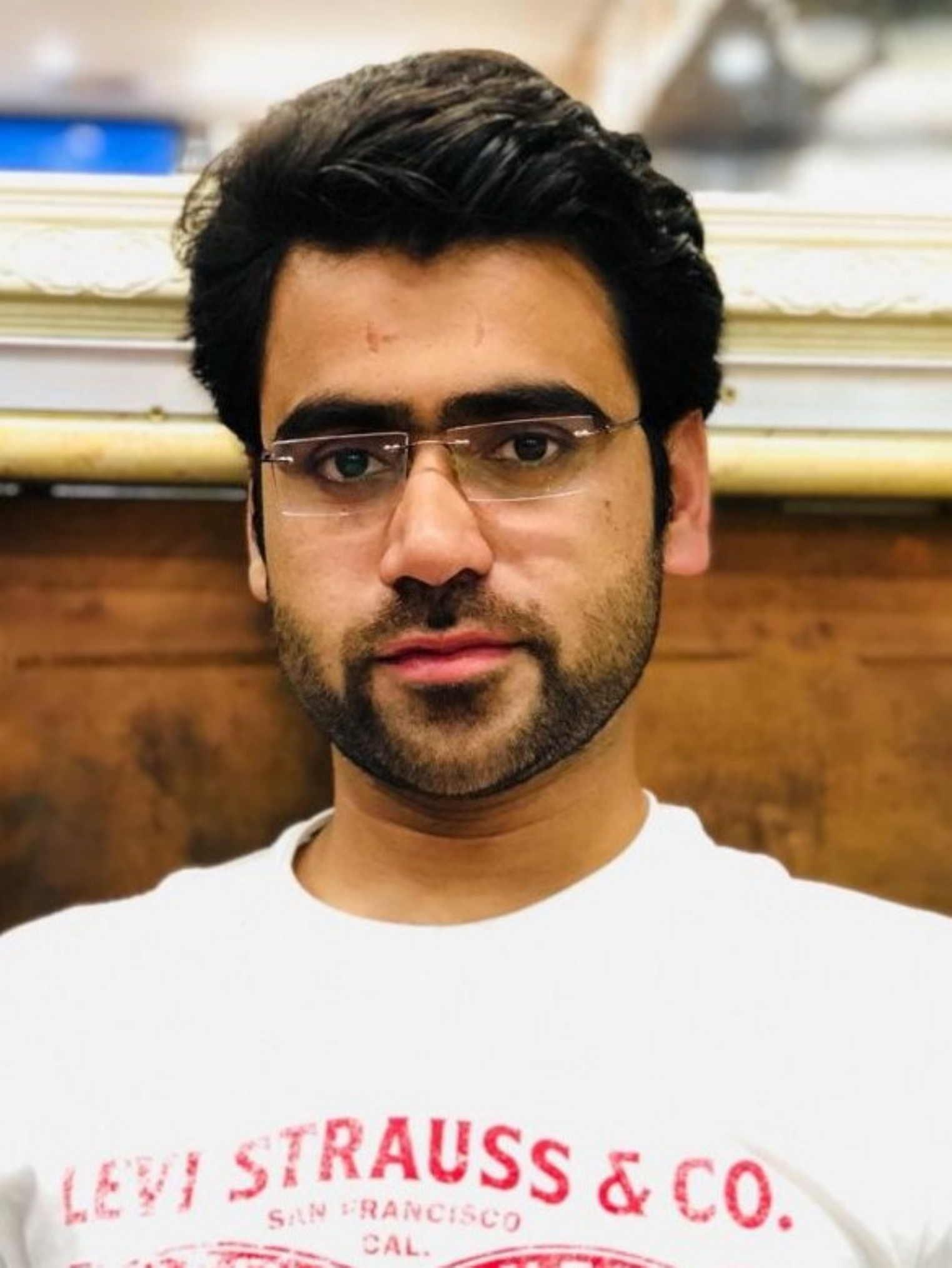}}]{Zafaryab Rasool}
received the BS and MS degrees from Aligarh Muslim University and Jamia Millia Islamia, India in 2013 and 2016 respectively. Currently, he is pursuing PhD with the Department of Computer Science and Software Engineering, Swinburne University of Technology, Australia.
His research interests include database systems, data mining and more specifically focus on clustering techniques
and algorithms.
\end{IEEEbiography}
\vspace{-24pt}
\begin{IEEEbiography}[{\includegraphics[width=1in,height=1.25in,clip,keepaspectratio]{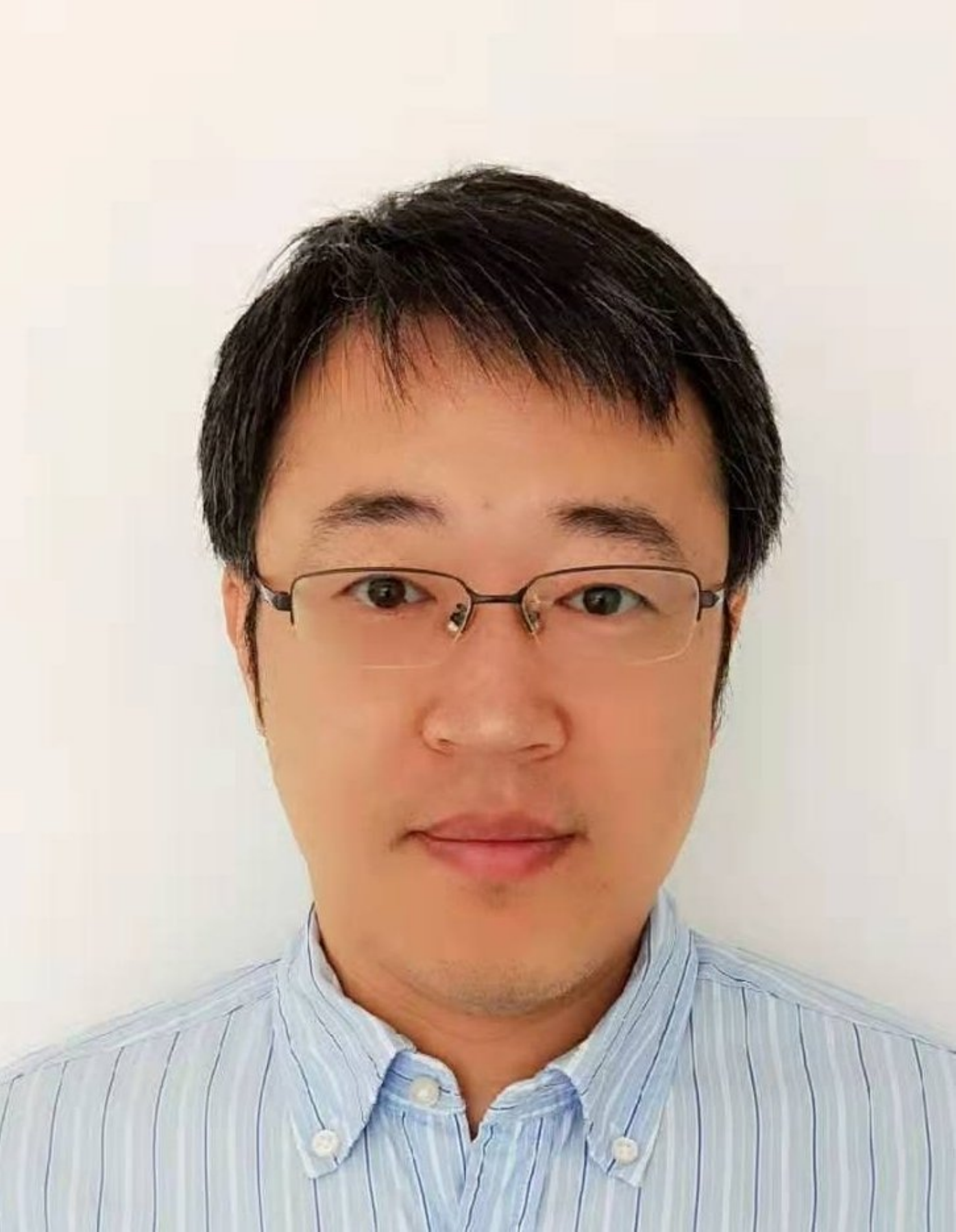}}]{Rui Zhou}
received the BS and MS degrees
from Northeastern University, China, in 2004 and
2006, respectively, and the PhD degree from the
Swinburne University of Technology, Australia, in
2010. He is currently a Lecturer in the in the Department of Computer Science and Software Engineering, Swinburne
University of Technology, Australia. His research
interests include database systems, data
mining, algorithms and big data. He is a member of the IEEE.
\end{IEEEbiography}
\vspace{-24pt}
\begin{IEEEbiography}[{\includegraphics[width=1in,height=1.25in,clip,keepaspectratio]{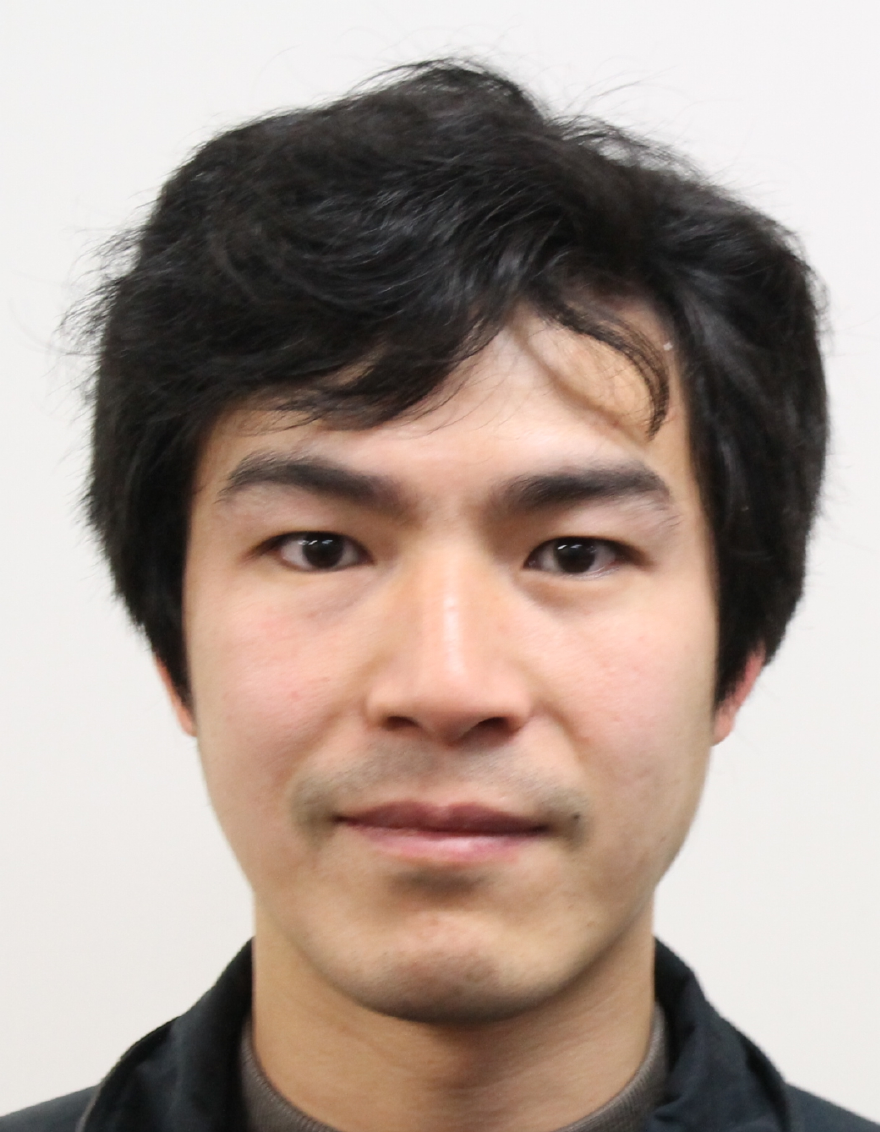}}]{Lu Chen}
received the BS degree
from Nanjing University of Technology, China, in 2008 and MS and PhD from Swinburne University of Technology, Australia, in 2014 and 2018, respectively. He is currently a Research fellow in the Department of Computer Science and Software Engineering, Swinburne
University of Technology, Australia. His research
interests include graph data management, algorithms and social network.
\end{IEEEbiography}
\vspace{-24pt}
\begin{IEEEbiography}[{\includegraphics[width=1in,height=1.25in,clip,keepaspectratio]{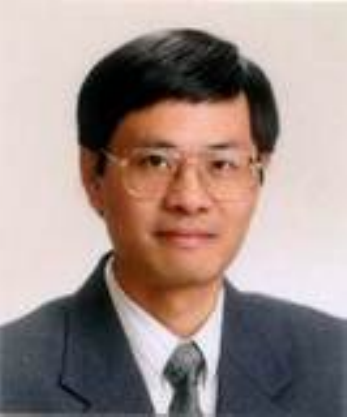}}]{Chengfei Liu} received the BS, MS and PhD degrees in Computer Science from Nanjing University, China in 1983, 1985 and 1988, respectively. He is a Professor in the Department of Computer Science and Software Engineering, Swinburne University of Technology, Australia. His current research interests include graph data management over large networks, keyword search on structured data, query processing and refinement for advanced database applications, and data-centric workflows. 
He is a member of the IEEE and ACM.
\end{IEEEbiography}
\vspace{-24pt}
\begin{IEEEbiography}[{\includegraphics[width=1in,height=1.25in,clip,keepaspectratio]{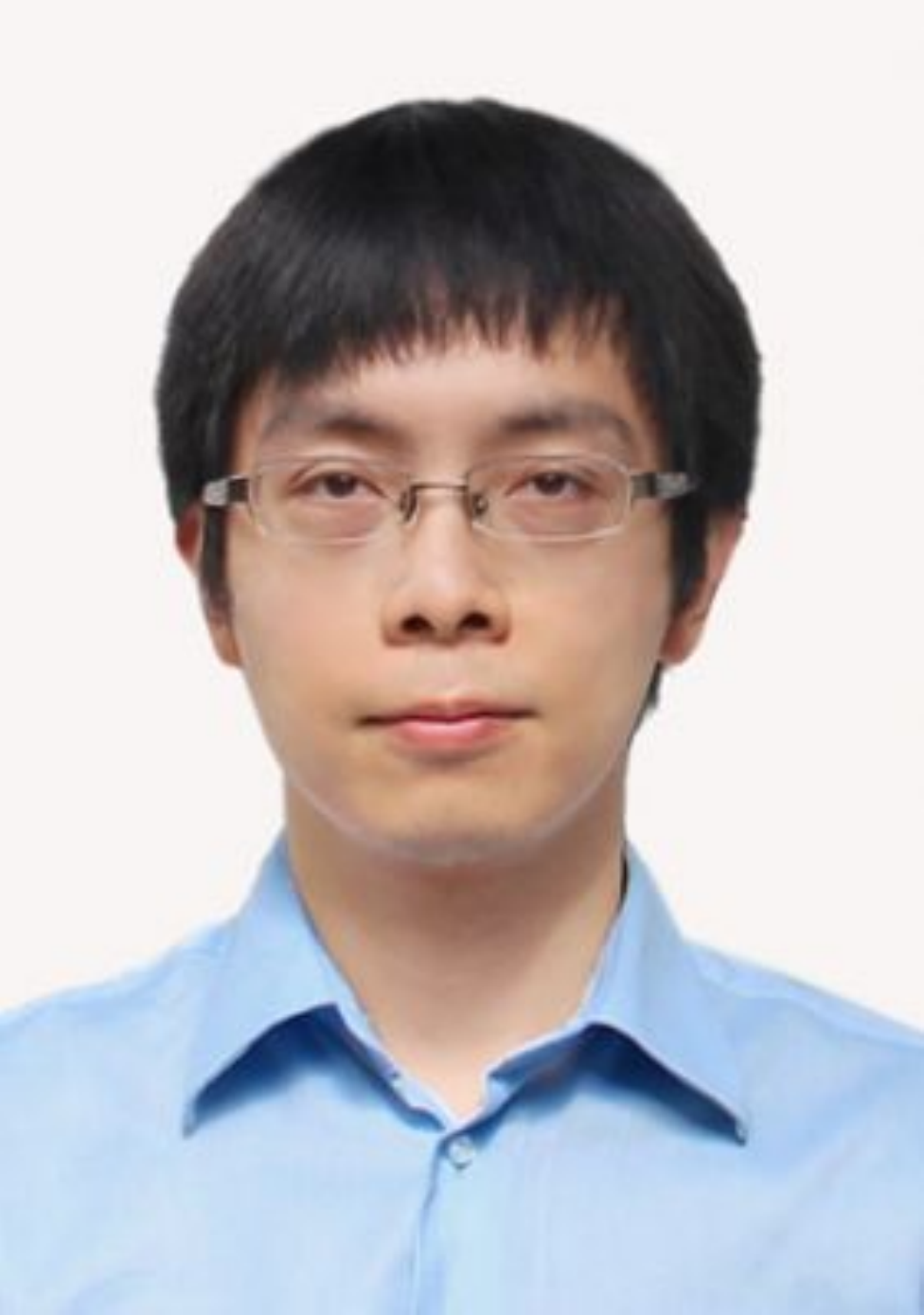}}]{Jiajie Xu}
received the MS degree from the University
of Queensland, Australia, in 2006 and
the PhD degree from the Swinburne University
of Technology, Australia, in 2011. He is currently
an Associate Professor with the School of Computer
Science and Technology, Soochow University,
China. His research interests include spatiotemporal
database systems, big data analytics
and mobile computing.
\end{IEEEbiography}
\vfill

\end{document}